# A Survey and Tutorial of EEG-Based Brain Monitoring for Driver State Analysis

Ce Zhang[1], Azim Eskandarian[2]

*Abstract*—Drivers' cognitive and physiological states affect their ability to control their vehicles. Thus, these driver states are important to the safety of automobiles. The design of advanced driver assistance systems (ADAS) or autonomous vehicles will depend on their ability to interact effectively with the driver. A deeper understanding of the driver state is, therefore, paramount. EEG is proven to be one of the most effective methods for driver state monitoring and human error detection. This paper discusses EEG-based driver state detection systems and their corresponding analysis algorithms over the last three decades. First, the commonly used EEG system setup for driver state studies is introduced. Then, the EEG signal preprocessing, feature extraction, and classification algorithms for driver state detection are reviewed. Finally, EEG-based driver state monitoring research is reviewed in-depth, and its future development is discussed. It is concluded that the current EEG-based driver state monitoring algorithms are promising for safety applications. However, many improvements are still required in EEG artifact reduction, real-time processing, and between-subject classification accuracy.

## I. Introduction

The NHTSA reports that human errors were responsible for 94% of the fatal crashes in 2016 [1]. The driver state and physiological condition affect his/her ability to control the vehicle. Therefore, by monitoring the driver state, one can predict anomalies or the potential for error and hence devise methods to prevent the consequences of human error. There are two solutions for human error minimization: fully automated vehicles (SAE Level 5) and driver monitoring systems (DMS). The first option eliminates the problem by totally removing the driver from controlling the car. The second option aims to monitor the driver and the driving task and assist the driver (or both the driver and the vehicle) in overcoming potential errors or hazards. Even though level 5 automated vehicle-related research shows promising results, fully automated vehicles will not be ready for the road in the foreseen future [2, 3]. Therefore, DMS is crucial to reducing human errors at present. DMS belongs to the family of advanced driving assistance systems (ADAS). Nevertheless, unlike other ADAS functions measuring vehicle performance, such as lane detection/control and traction control, DMS directly measures the driver's state and behavior during driving. Since the majority of human errors are driver distraction, inattention, fatigue, and drowsiness, most DMS are designed for driver cognitive state surveillance and driver attention improvement [4].

The DMS can be designed for application or for research purposes. The application systems are designed for on-market vehicles. Hence, they are compact and economical but have limited driver state detection ability. Application-based DMS were found first in Toyota and Lexus in 2007 [5]. They use a camera to monitor the driver's face and eyes, as shown in Fig 1a. Once the driver's eyes are closed and the face is not facing forward, the DMS starts to warn the driver and decrease vehicle speed. A similar device was also designed by BMW (Fig 1b) named as "Driver Attention Camera" [6]. The BMW monitoring system can detect whether the driver's eyes are focused on the road. In addition to eye and face tracking, a pressure sensor was developed and applied on Tesla to force the driver to put his/her hands on the wheel while driving [7]. Currently, most application-based DMS apply simple algorithms that can only detect drivers' physical behaviors instead of drivers' cognitive states, such as "daydreaming" and distraction during driving. The DMS designed for research are more complex and expensive but can detect different types of driver states. Research-based DMS employ different sensors for surveilling and analyzing driver behavior. Popular sensors include electroencephalography (EEG), eye tracker, body motion tracker, handgrip sensor, electrocardiography (ECG), and electromyography (MEG). These sensors are complex to analyze and expensive but can provide more comprehensive information about the driver's driving condition and driver state.

Among all the research-based DMS sensors, EEG is one of the most effective driver state monitoring devices. The main reasons are that i) EEG collects human brain signals, which directly measure drivers' cognitive states and thoughts, and ii) EEG signal temporal resolution is high, which can provide more neural-related activity from the driver [8]. Therefore, for research-based DMS, researchers

Ce Zhang is with Virginia Tech Mechanical Engineering Department ASIM LAB (email: zce@vt.edu).

Azim Eskandarian is the Head of the Virginia Tech Mechanical Engineering Department, Nicholas Rebecca Des Chaps Professor, and the Director of the ASIM Lab (email: eskandarian@vt.edu).

usually apply EEG signals not only for driver state but also as an evaluation baseline, such as finding correlations between the EEG results and eye gaze signals or correlations between the EEG results and ECG signals in driver's state detection [9, 10].

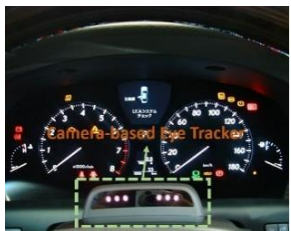 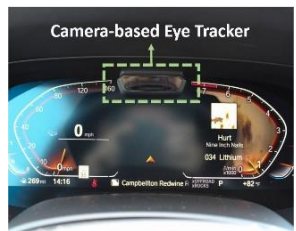

Fig 1a.  Sample Driver Monitoring System (Lexus)  
Fig 1b.  Sample Eye Tracking Sensor (BMW)

The history of EEG-based driver condition studies can be traced back to the early 1980s. M. Lemke proved that driver vigilance decreases during long-term driving through EEG measurement [11]. K. Idogawa et al. compared a professional driver's EEG signal with a regular driver [12]. They found that brain waves move from the beta to alpha domains in monotonous tasks such as driving. However, due to computational limitations and a lack of pattern recognition algorithm support, EEG-based driver studies did not become popular until the start of the 21st century. S. Lal et al. developed a LabVIEW program to categorize driver fatigue levels based on different brain rhythm spectra [13]. In 2013, D. Gohring et al. collected and processed driver EEG signals that successfully controlled a vehicle through the brain at low speed [14]. Presently, EEG-based driver state studies have become systematic in both the study of interest areas and data analysis. The study areas can be categorized as driver distraction/inattention, fatigue/drowsiness, and intention study [15, 16]. The objective of the driver distraction/inattention study is to evaluate and classify driver distraction levels under different types of nondriving-related tasks (secondary tasks). Researchers instruct testing subjects to conduct one or multiple secondary tasks, such as texting, talking, or watching videos while driving [17]. Additionally, the EEG device collects the subject brain signal. Both EEG signals and vehicle performance are used as evaluation references. The distraction detection accuracy varies greatly among every literature because of different experimental environments, subjects, and processing algorithms. The driver fatigue/drowsiness experiments study and classify driver fatigue levels during driving. To achieve a drowsiness state, testing subjects are usually required to drive for an extended period (70-90 minutes) at a specific time (midnight or afternoon). "Vehicle lane departure", reaction time when facing emergency conditions, and EEG signal results are usually applied for drowsiness level evaluation [18]. The detection accuracy also varies among different studies, but most research results indicate that the higher the drowsiness level is, the more steering movement exists [19-21]. Driver intention studies are mostly focused on driver brake intention studies. Driver brake intention experiments can be categorized into two types. The first type of braking study aims to find driver reaction time among different ages, sexes, or driving conditions [22-25]. The second type investigates the time difference between the brain brake intention and actual driver brake action [26-29]. In the second type of experiments, numerous braking warning systems can be designed based on either improving brake intention or minimizing the abovementioned time difference. Since driver state analysis contains fatigue and distraction studies, most experiments are conducted under a lab environment with a driving simulator, as shown in Fig 2. However, several experiments are conducted using a real vehicle under the real-world environment, such as the brake intention experiment [29]. Fig 3 presents a summary of several representative studies in EEG-based driver state studies in recent years. The EEG data analysis method is a data-driven study predicting driver behavior based on previously collected data through machine learning techniques. A flowchart of a typical EEG-based driver behavior data analysis is presented in Fig 4. The signal preprocessing step is designed for noise reduction and artifact removal. The feature extraction extracts spatial, temporal, and frequency features for model development. The classification step builds a mathematical model based on the training data. The details of these methods are discussed in the remaining sections of this article.

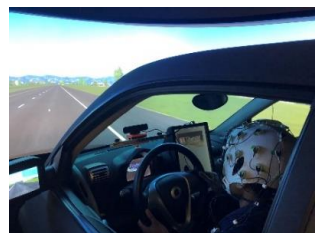 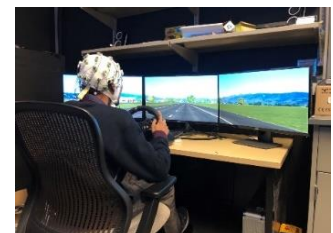

Fig 2a.  Vehicle-based simulator  
Fig 2b.  Desktop-based simulator

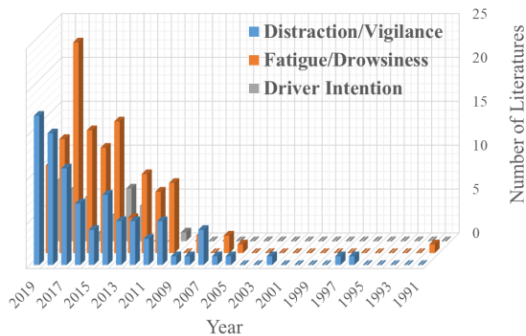

Fig 3.  EEG-based Driver State Study Representative Literature

As illustrated above, EEG behavior on driver condition has been studied for over three decades, and numerous data processing and model development techniques exist. The studies for EEG features and classification have also evolved from simple time or frequency domain analysis to statistical analysis based on big data. The objective of this paper is to document and review the majority of popular experiments and data processing methods for EEG behavior in driver condition-related research. In section II, an overview of the EEG signal collection apparatus and methods are briefly illustrated. In sections III to V, the EEG data preprocessing, feature extraction, and data classification algorithms are explained in detail. In section VI, we demonstrate and summarize EEG data processing research on driver conditions and predict their future development.

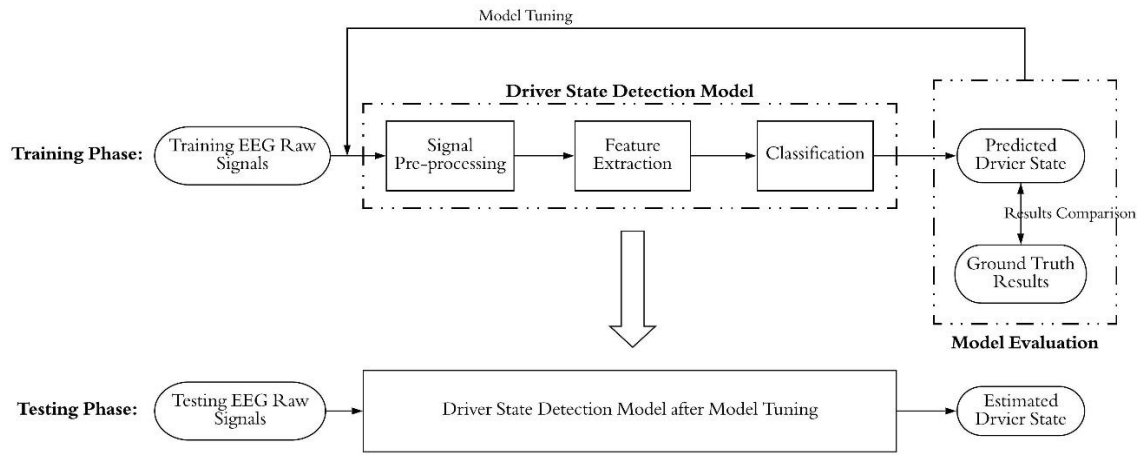

Fig 4. EEG-based Driver State Study Workflow

## II. EEG Signal Collection Methods

A typical EEG signal collection system is composed of a signal acquisition cap, an amplifier, and a data storage device, as shown in Fig 5. After data collection, corresponding data analysis software is used for data postprocessing. For the amplifier, most EEG-based driver state studies employ a commercially designed amplifier such as gUSBamp [30]. For data storage devices, laptops, or microprocessors such as Raspberry Pi are popular options. For data analysis software, there exist either commercial software such as for BrainVision Analyzer-related products or open-source data reading and analysis toolboxes such as EEGLAB and OpenBCI [31-33]. The abovementioned devices are standard in EEG driver state-related research, and the results are not significantly affected by alternative selections. However, for signal acquisition caps, different types of devices can cause distinct experimental results. The remaining paragraphs in this section introduce the EEG signal acquisition system categorization and the corresponding properties.

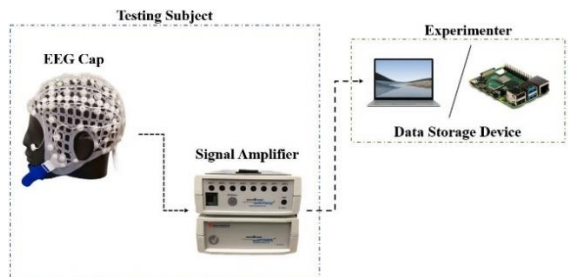

Fig 5. EEG Signal Collection System

Most EEG signal acquisition systems applied in driver state studies can be categorized as dry electrode vs. wet electrode and wire communication vs. wireless communication. In addition, several researchers use a remote headband for EEG signal collection. The next paragraphs explain the details of every categorization.

### A. Wet Electrode vs. Dry Electrode

Wet and dry electrode EEG caps are shown in Fig 6. The wet electrode is made of silver or silver chloride material, which are commonly used in lab environments. The dry electrode uses stainless steel as a conductor to transmit the microvolt signals from the brain surface to the EEG amplifier. Compared with dry electrodes, electrolytic gel must be injected between the electrode and subject brain surface to increase the signal impedance and improve accuracy. Hence, the experimental procedure for the wet electrode EEG cap is more complicated than that for the dry electrode. Moreover, both the EEG cap and subject's head need to be cleaned after every experiment, which is time-consuming and inconvenient. However, according to K. Mathewson et al. [34], wet EEG electrodes exhibit less noise than dry electrodes, and the root mean square (RMS) results for brain event detection are lower. Therefore, from the perspective of driver study, the wet electrodes fit simulator-based research better, but the dry electrodes are more convenient under real-world driving conditions.

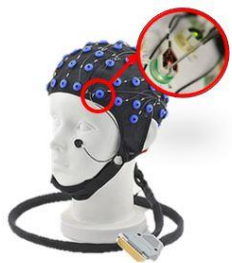 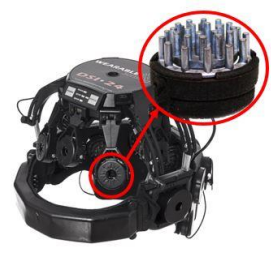

Fig 6a. Wet Electrode  Fig 6b. Dry Electrode

## B. Wire Connection vs. Wireless Connection

The wired and wireless connections mentioned in this paragraph are discussed in the scope of connectivity between the EEG cap and the amplifier. The wired connection-based EEG cap has redundant wires, which are easily broken by the subject during driving events. Besides, cable sway during driving can cause motion artifacts that affect the EEG collection accuracy [35]. For EEG wireless connections, the most popular connection method is through Wi-Fi. Wireless connections are more convenient and compact because they eliminate redundant wires. However, the main drawback of wireless EEG devices, based on the research results of A. Torok, et al., are the electrical noises during wireless transmission [36]. When external electrical effects exist, the EEG signals collected by the wireless method are contaminated. The other drawback of wireless connections in driver state studies is the loss of connections. Presently, communication stability and loss of information under real-world driving conditions are still the most important challenges in connected vehicles [37]. Hence, using a wireless EEG cap could experience loss of information problems. Therefore, at present, a wired connected EEG signal cap for driver state study is still a better choice due to the accuracy and completeness of signal information.

## C. Traditional Electrode vs. Headband

Currently, most traditional electrodes and headband positions are based on the 10-20 international system electrode positioning standard that was adopted around 1958 [38]. However, the traditional EEG cap can collect signals among the whole brain region, while the headband can only collect signals from the forehead, as shown in Fig 7. Since the traditional EEG cap collects multiple channel signals, it can detect more brain activities. Typical selection about the number of channels for driver state study is 21, 32, and 64 channels. Among these channels, the C3, Cz, and C4 electrodes are most important because they measure the area that is in charge of the driver's thought and motor movement [39-41]. Furthermore, after multidimensional data are obtained, several advanced data processing techniques can be applied for analysis, which is illustrated in the following sections. However, multidimensional data analysis is a double-edged sword. Multidimensional data processing requires advanced signal processing knowledge, and the computational load can be extremely high. In contrast to traditional electrodes, the headband only collects four channels from the human forehead, and the price is lower compared with most EEG signal acquisition systems. R. Foong employed a muse headband for driver vigilance detection [42, 43]. Although the headband can detect brain activities, the device accuracy and robustness still need to be verified.

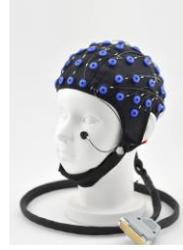

Fig 7a. Traditional Electrode Cap

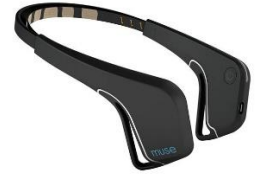

Fig 7b. Headband

## III. EEG Signal Preprocessing

The objective of signal preprocessing is artifact removal. According to M. Sazagar and M. Young, there are two types of EEG artifacts, non-physiological and physiological [44]. The non-physiological artifacts are mainly caused by the EEG amplifier, external noises, and electrodes. Usually, in the context of EEG-based driver state studies, non-physiological artifacts are removed by a linear bandpass filter with bandwidth ranges from 1 to 50 Hz. Physiological artifacts are generated by testing subjects and can be categorized as ocular, muscle, and cardiac artifacts [45]. In the context of driver state studies, muscle and ocular artifacts commonly occur and must be removed. Unlike non-physiological behavior, the abovementioned artifacts can be considered a measure of subject behavior, and the majority of them exist in a similar bandwidth with the desired EEG signals. Thus, the identification and removal of these artifacts require a more complicated data analysis algorithm. Table I tabulates a summary of several popular artifact removal algorithms that are employed for driver state analysis applications. In the remaining paragraphs, these algorithms and their corresponding alternative forms are illustrated in detail.

### A. Independent Component Analysis

ICA is a blind source separation algorithm for multivariate signal processing. Two major assumptions for ICA are as follows: first, the mixture signals are composed of several statistically independent components; second, the relationship between the mixture signals and every independent component is linear. Thus, the ICA equation is

$$X_{[n \times k]} = W_{[n \times m]} * S_{[m \times k]} \quad (1)$$
$$S_{[n \times k]} = A_{[m \times n]} * X_{[n \times k]} \quad (2)$$

where Equation (1) is the reconstruction formula and Equation (2) is the decomposition function. In these equations, $X$ is the collected multichannel EEG signal matrix with $n$ channels and $k$ samples, S is the independent component matrix with user-defined $m$ components, W is the transformation matrix, and $A$ is the pseudoinverse of the $W$ matrix. ICA is an unsupervised learning process, and the transformation matrix $W$ is acquired through multiple iterations. R. Nuno et al. proved that the application of the ICA algorithm for ocular artifact removal in EEG signal processing is feasible [46]. In their algorithm, eye activity and brain activities are separated by amplitude kurtosis. Even though the ICA algorithm can separate ocular artifacts effectively, the algorithm's computational load is high, and the detection process is manual. Therefore, there exist several modified ICA algorithms to achieve faster processing speed or enable automatic detection.

*1) Improved Processing Speed*

The key reason for the high computational load of the ICA algorithm is a lack of prior knowledge input. As mentioned above, ICA is an unsupervised learning algorithm that requires the computer to estimate every weight factor through multiple iterations. Thus, employing EEG artifact prior knowledge is an effective method for decreasing the computational load. B. Peters modified the ICA by adding reference signals [47]. With the help of reference signals, only desired components are analyzed, which improves the processing speed. M. Akhtar et al. proposed a spatially constrained ICA (SCICA) [48]. In this algorithm, only artifact-related components are extracted, which decreases the computational load and achieves a high extraction rate as well.

*2) Automatic Detection*

To solve the automated detection issue, C. Ahlstrom proposed an automated artifact handling algorithm (ARTE) to automatically detect and remove artifacts through ICA [49]. In their paper, wavelet decomposition and hierarchical clustering methods are combined with ICA. The use of wavelet decomposition extracts 2-second segments, and the application for hierarchical clustering automatically separates artifacts from EEG signals. According to the comparison results with other state-of-the-art algorithms, the ARTE outperforms other algorithms and is suitable for artifact removal for EEG signals collected in a naturalistic environment. I. Winkler et al. also proposed an automatic ICA artifact detection algorithm [50]. During the ICA, they take the temporal correlations, frequency and spatial features as detection factors.

In summary, the present modified ICA algorithms can automatically detect and extract ocular artifacts, and the computational speed is improved by adding prior artifact knowledge as input. Nevertheless, since ICA is based on estimation, ocular artifacts cannot be completely removed. In the future, how to maximally remove all physiological artifacts still needs to be resolved.

### B. Canonical Correlation Analysis (CCA)

CCA is also a blind source separation algorithm. However, the goal of CCA is to seek the maximum correlation between two multivariate datasets. More specifically, assume $X$ and $Y$ are two collections of the dataset. The CCA algorithm attempts to find vectors $a_x$ and $a_y$ such that

$$\max_{a_x, a_y} \rho(a_x X, a_y Y) = \frac{E[a_x X a_y Y]}{\sqrt{E[(a_x X)^2]E[(a_y Y)^2]}} \quad (3)$$

where $\rho$ is the correlation factor between $a_x X$ and $a_y Y$. By taking the derivative of Equation (3) with respect to $a_x$ and $a_y$, the maximum correlation factor yields

$$\begin{cases} C_{xx}^{-1} C_{xy} C_{yy}^{-1} C_{yx} a_x = \rho^2 a_x \\ C_{yy}^{-1} C_{yx} C_{xx}^{-1} C_{xy} a_y = \rho^2 a_y \end{cases} \quad (4)$$

where $C_{xx}$ and $C_{yy}$ are the autocovariance of $X$ and $Y$, and $C_{xy}$ and $C_{yx}$ are the cross-covariance between $X$ and $Y$. The CCA algorithm was first applied for EEG muscle artifact removal by S.V. Huffel and her colleagues [51]. In that paper, two data matrices $X(t)$ and $Y(t)$ were selected for CCA, where $X(t)$ is the time-series brain signal and $Y(t)$ was a one-sample delayed version of $X(t)$. According to their signal-to-noise ratio comparison test, the CCA algorithm outperforms ICA and the low-pass filter method for muscle artifact removal. However, the conventional CCA algorithm requires manual labeling for muscle artifacts. Thus, several improved CCA algorithms have been aimed at achieving automatic detection. In addition, since CCA requires less computation load, there are some algorithms to modify the CCA algorithm and implement it for real-time artifact detection and artifact removal.

*1) Automatic Detection*

To achieve automatic detection, prior knowledge about EEG artifacts is required. J. AS et al. improved the conventional CCA method by combining EEG spectral knowledge for automatic detection [52]. They employed spectral slope analysis for artifact detection and set a threshold for the correlation coefficient to remove several components. The experimental results proved that the modified CCA algorithm could remove high-frequency muscle contamination. M. Jatoi combined empirical mode decomposition (EMD) and CCA for automatic eye blink artifact detection [53]. The EMD was used as a signal decomposition algorithm to detect the eye blink template. After that, CCA was applied to remove those artifacts. Experimental results indicated that the EMD-CCA algorithm can be easily adjusted and applied to every EEG electrode.

*2) Real-Time Processing*

Compared with ICA, CCA requires a lower computational load. Hence, it is easy to implement for real-time artifact removal processing. P. Wang et al. applied CCA for real-time muscle artifact removal [54]. Based on their results, the CCA for muscle artifact removal outperforms the ICA algorithm. C.T. Lin and Y.K. Wang proposed a real-time artifact removal algorithm based on CCA [55]. They applied the CCA algorithm to decompose the EEG signal and used a Gaussian mixture model to classify artifacts. This algorithm is helpful in driver state studies because it can detect artifacts commonly occurring during driving, such as eye blinks and head/body movement.

In general, the CCA algorithm can accurately remove muscle artifacts and requires a lower computational load. In addition, there are built-in libraries such as "canoncorr" in MATLAB and open-source packages in Python ("scikit-learn") [56, 57]. The implementation of CCA for artifact removal in EEG-based driver state studies is convenient.

*C. Wavelet Transform (WT)*

WT can be considered as an alternate form of Fourier transform. Instead of transforming every piece of the signal into a sine wave, the WT applies a specific waveform to decompose signals. WT can be categorized as a continuous and discrete wavelet transform (CWT/DWT). CWT is a signal processing technique for nonstationary signal time-frequency analysis, and DWT is commonly applied for signal denoising and artifact removal [58]. For DWT analysis, the input signal is decomposed into detail and approximate information with a high-pass and a low-pass filter, respectively, and the formula is

$$\begin{cases} y_{low}[n] = x[n] * g[n] \\ y_{high}[n] = x[n] * h[n] \end{cases} \quad (5)$$

where $g[n]$ is the low-pass filter and $h[n]$ is the high-pass filter. For multilevel decomposition, the approximate information is selected for next level decomposition, and the detailed information is expressed as the level coefficient, as shown in Fig 8. After wavelet decomposition, a threshold is applied to discard signals with artifacts. This algorithm brings up two concerns: first, how to select the mother wavelet and the number of decomposition levels; second, how to avoid overfiltering.

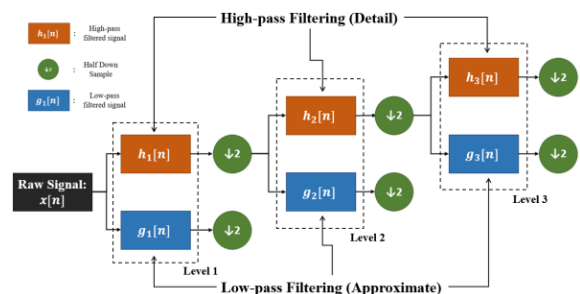

Fig 8. DWT 3 Level Decomposition Flowchart

*1) Decomposition Level and Waveform Selection*

For DWT, the choice of waveform and decomposition level is critical for artifact removal effectiveness. Unfortunately, there is no exact answer to the number of decomposition levels and waveform choices because human subjects and experimental conditions are varied, which strongly affects artifact behavior. Thus, these values need to be determined according to specific experiments. S. Khatun et al. compared different wavelet performances for ocular artifact removal [59]. The "Symlet", "Haar", "Coiflet", and "biorthogonal wavelets", as shown in Fig 9, are used for wavelet decomposition, and the decomposition level is set to eight. The results indicate that coif3 and bior4.4 are more effective. In a V. Krishmnaveni ocular artifact removal study, she found that a 6-level Haar wavelet transform outperforms the other state-of-the-art DWT algorithm [60]. Even though the number of levels and wavelets are hard to determine, it is clear that the decomposition level needs to achieve the desired frequency range and the mother wavelet needs to be similar to the signal.

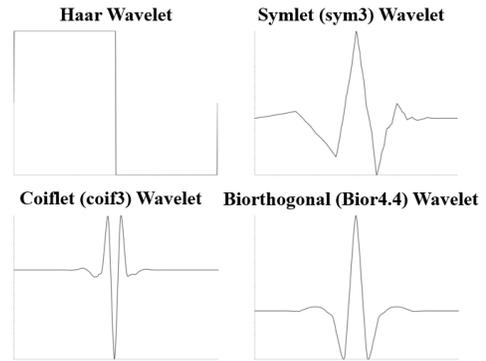

Fig 9. Common Wavelet Functions for the EEG DWT Algorithm

*2) Overfiltering Avoidance*

The ocular artifact exhibits spectral properties similar to those of the EEG signals. When dealing with ocular artifact removal, the DWT algorithm has the potential to remove not only EOG artifacts but also useful EEG information. Therefore, to avoid over-filtering, the DWT algorithm is often combined with other source separation algorithms. ICA is a popular choice because it helps with removing the most useful EEG information from the artifact components. The details about the ICA and DWT combination algorithms are explained in the hybrid detection section.

*D. Regression Analysis*

Regression analysis is a statistical method for exploring the relationship between multiple variables of interest. In the context of EEG artifact removal, the variables of interest are EEG signals and artifacts. The assumption for regression analysis is that the measured EEG signal is composed of pure EEG and artifacts. as shown in Equation (6)

$$EEG_{measure} = EEG_{correct} + p * EOG \quad (6)$$

where $EEG_{measure}$ is the collected EEG signal, $EEG_{correct}$ is the ground-truth pure EEG signal, $EOG$ is the ocular signal, and $p$ is the weighted factor. The objective of regression analysis is to estimate the weighted factors so that the estimated "pure EEG signals" are close to the ground truth. Unlike blind source separation, regression analysis requires one or more reference channels. The regression technique for removing the EOG artifact can be achieved by either time or frequency domain analysis [61, 62]. Both domains could achieve good results, but time-domain analysis can achieve better EOG artifact removal performance with an appropriate choice of an adaptive filter [45].

The advantages of regression analysis are easy modeling, and the computational load is extremely low. Since there exist one or multiple reference channels, EEG artifacts can be easily extracted based on those references. However, in some EEG experiments, reference channels are not available. Even though the reference channels are obtained, the EEG signals are usually also contained in those channels. Therefore, for present studies, blind source separation algorithms such as ICA, CCA, and wavelet transform are more popular than regression analysis.

*E. Hybrid Detection Methods*

All the algorithms have advantages and drawbacks. Therefore, combining those algorithms to overcome those drawbacks and mutual beneficiation is the trend for the current artifact removal study. Z. Tian et al. conducted a comprehensive review of several hybrid detection methods [45]. In this section, we briefly introduce the ICA and DWT combination method, regression analysis and ICA combination method because they are commonly applied in EEG-based driver state studies.

*1) Combination between ICA and DWT*

Both ICA and DWT algorithms can cause over-filtering issues [63]. After typical ICA processing, the extracted artifact components usually still contain the remaining EEG signals. Thus, removing the entire component may cause a loss of EEG information. A similar issue also occurs in DWT processing because some artifacts, such as ocular artifacts, share similar spectral properties with the desired EEG signals. Therefore, some studies attempted to combine ICA and DWT to determine whether these two algorithms can mutually benefit each other.

P. Funk combined the ICA and DWT to automatically remove EEG artifacts [49]. In their paper, wavelet decomposition was applied to the recorded EEG signals for several 2-second segment signals. After wavelet decomposition, the ICA algorithm was applied to remove EEG artifacts. According to their results, the ICA and DWT combination algorithm exhibited better artifact removal effectiveness. M. Issa et al. also combined ICA with wavelet decomposition to prevent over-filtering issues [64]. They conducted ICA first and used DWT to decompose extracted artifact components. With this improved algorithm, the artifact cleaning results outperformed other state-of-art artifact removal algorithms.

However, the computational load for the ICA and DWT hybrid detection algorithm is high. In EEG-based driver state studies, the ability for real-time processing is required. Therefore, improving this hybrid detection algorithm processing speed is an interesting but challenging study for the future.

*2) Combination between Regression Analysis and ICA*

Similar to the DWT and ICA combination algorithm, hybrid detection with regression analysis and the ICA method also tries to avoid over-filtering issues. As mentioned above, the artifact components after ICA decomposition may still contain neural-related activities. For regression analysis, the selected reference channels may also provide useful EEG information. Therefore, combining the ICA and regression analysis can avoid eliminating too much desired neural information.

M. Mannan proposed an automatic artifact identification and removal algorithm based on ICA and regression analysis [65]. ICA was first applied for component decomposition, and then regression analysis was used for artifactual components. Since they assumed that artifactual components contain few neural activities, regression analysis-based artifact removal can minimize the error of over-filtering.

TABLE I
EEG ARTIFACT REMOVAL ALGORITHM METHODOLOGY SUMMARY

| | | EEG Artifact Removal Methods | | | |
|---|---|---|---|---|---|
| Algorithm | Algorithm Summary | Related Physiological Artifact | Advantages | Drawbacks | Related Literatures |
| Independent Component Analysis (ICA) | Blind source separation extracting components are orthogonal to each other | Ocular | • Effectively removed artifacts Unsupervised learning | • Prior knowledge about the artifacts;<br>• High Computation Load | [46-50] |
| Canonical Correlation Analysis (CCA) | Blind source separation finding maximum correlation | Muscle | • Low computation load | • Prior knowledge about the artifacts | [51-57] |
| Wavelet Transform (WT) | Decompose signals, setup threshold, and reconstruction | Ocular | • Non-stationary signal analysis; | • Trivial Decomposition Level and Wavelet Selection | [58-60] |
| Regression Analysis | Finding artifacts based on the reference channels | Ocular | • Easy modeling;<br>• Low computation load | • Requiem one or multiple reference channels | [61-62], [45] |
| Hybrid Method | Combination of multiple algorithms to achieve complementary advantages | Ocular + Muscle | • Comprehensive artifact removal;<br>• Mutual beneficiation | • High computation time;<br>• Complex modeling | [49], [64-65] |

## IV. EEG FEATURE EXTRACTION METHODS

Feature extraction is a term in machine learning that obtains desired information from redundant noisy signals [66]. Since EEG signals are nonstationary and usually collected in high dimensions, feature extraction is necessary for filtering and dimension reduction. In an EEG-based driver state study, feature extraction algorithms based on temporal, frequency, and spatial domains are employed to obtain driver condition features based on prior known EEG features. In this section, we introduce common EEG features applied in driver state studies and then illustrate corresponding feature extraction algorithms.

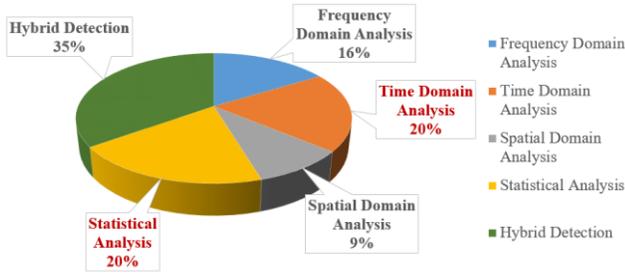

Fig 10. Literature Distributions for Common Feature Extraction Algorithms

### A. Common Driver-Related EEG Features

In this section, the event-related potentials (ERPs) and the event-related desynchronization/synchronization features in the temporal domain (ERD/ERS), EEG rhythms in the frequency domain, and brain lobe locations in the spatial domain are illustrated.

#### 1) ERP and ERD/ERS

ERP behavior is a small electric voltage change that can be measured by EEG in response to motor or cognitive event stimuli. Since EEG collected signals are contaminated by artifacts, averaging over multiple experimental trials is necessary to observe clear ERP results, as shown below.

$$x_{clear}(t) = \frac{1}{N} * \sum_{m=1}^{N} x_{raw}(t,m) \quad (7)$$

where $N$ is the total number of trials, $x_{raw}(t,m)$ is the $m^{th}$ trial signal, and $x_{clear}(t)$ is the averaged ERP result. Different waveforms exhibited in ERP behavior represent different brain events. The P300 waveform is one of the most popular waveforms for studying driver inattention and driver fatigue. Detailed descriptions of P300 and other common waveforms are documented in V. K. Sinha's study [67]. The major benefit of the ERP is the ease of calculation and observation. However, the ERP feature is both time-locked and phase-locked to brain stimulus events.

According to G. Pfurtscheller's research, ongoing EEG can also be processed for brain cognitive or motor events [68]. Then, he introduced the ERD/ERS features. The calculation of ERD/ERS contains bandpass filtering, squaring sample amplitude (power samples), and averaging the power samples across all trials. The ERD/ERS behavior could be observed through both the time domain and spatial domain. The major benefit of the ERD/ERS feature is the non-phase-locked to the event.

#### 2) EEG Rhythms

EEG data are nonstationary signals composed of different waves. Table II is a summary of different brainwaves and their corresponding mental states [69]. For the driver cognitive states study, alpha (8-12 Hz) and beta (12.5-30 Hz) are the most interesting and perhaps the most relevant rhythms, and the waveforms are presented in Fig 11 and 12, respectively.

TABLE II
BRAIN RHYTHMS AND THEIR CORRESPONDING MENTAL STATES

| Brain Rhythm Type | Frequency Range (Hz) | Mental States and Conditions |
|---|---|---|
| Delta | 0.1-3 | Deep Sleep, non-REM Sleep*, Unconscious |
| Theta | 4-7 | Intuitive, Creative, Recall, Fantasy, Imaginary, Dream |
| Alpha | 8-12 | Relaxed but Not Drowsy, Tranquil, Conscious |
| Low Beta | 12-15 | Formerly SMR*, Relaxed yet Focused, Integrated |
| Midrange Beta | 16-20 | Thinking, Aware of Self & Surroundings |
| High Beta | 21-30 | Alertness, Agitation |
| Gamma | 30-100 | Motor Functions, Higher Mental Activity |

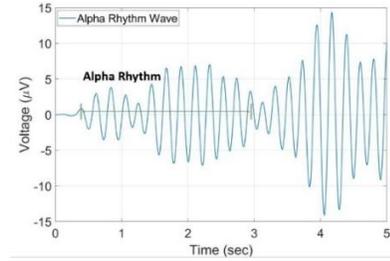

Fig 11. Alpha Waveform (Extracted from Experimental Data)

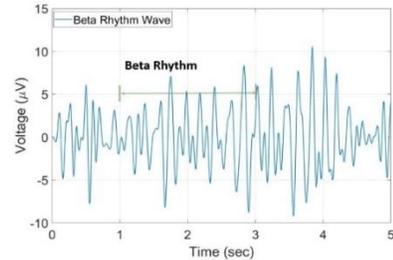

Fig 12. Beta Wave (Extracted from Experimental Data)

#### 3) Brain Lobes

The human brain (Fig 13) is composed of the forebrain, midbrain, and hindbrain [70]. The forebrain is one of the most important regions for EEG driver state study. The forebrain area can be categorized by four main lobes: the frontal lobe, parietal lobe, occipital lobe, and temporal lobe, as presented in Fig 14 [71]. The overall functions of each lobe are tabulated in Table III. In driver state studies, the frontal, parietal, and occipital lobes are the main areas of interest.

TABLE III
LOBES FUNCTIONS SUMMARY TABLE

| Brain Region | Main Functions |
|---|---|
| Frontal Lobe | Execution, Thinking, Planning |
| Motor Cortex | Motor Movement |
| Sensory Cortex | Sensation |
| Temporal Lobe | Memory, Language |
| Occipital Lobe | Vision |
| Parietal Lobe | Perception, Arithmetic |

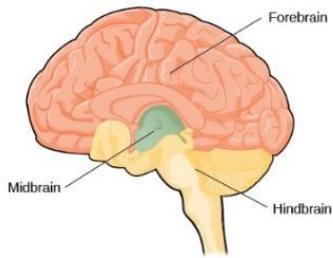

Fig 13. Human Brain Composition

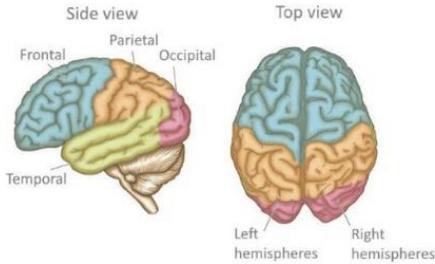

Fig. 14. Main Lobes for Human Forebrain Region

### B. EEG Feature Extraction Algorithm

The objective of the EEG feature extraction algorithms is to obtain uncovered features from temporal, frequency, and spatial domains for classification. According to the literature, the EEG-based driver state study feature extraction algorithm can be categorized as signal processing-based methods and statistical-based methods, as shown in Table IV. This chapter explains and evaluates every popular feature extraction algorithm tabulated in Table IV.

#### 1) Signal Processing-based Methods

Signal processing-based methods employ classical signal analysis techniques to extract features from temporal, frequency and spatial frequencies. Common spatial pattern (CSP) and spectral analysis algorithms are illustrated in this section.

#### a) Common Spatial Pattern (CSP)

Originally, the CSP algorithm was popular in human motor imagery analysis [72]. Currently, the CSP algorithm is widely employed in driver state studies because drivers' thoughts and cognition are similar to motor imagery behavior. The objective of the CSP algorithm is to estimate a transformation matrix so that the transformed EEG signal dimensions are reduced and the remaining signal variances can be distinguished between different classes. For instance, two-class multidimensional EEG signals are processed by the CSP algorithm. After processing, the EEG signal dimensions are reduced to two. Moreover, class A transformed signal variance is maximized in dimension 1 and minimized in dimension 2, while the class B transformed signal variance is the opposite, as shown in Fig 15a and 15b. To obtain the CSP transformation matrix, the eigenvalues and eigenvectors of the covariance matrices from each class of EEG signals are required, which is explained in detail in [72]. The application of the CSP algorithm for driver state study is usually on driver intention and cognitive load analysis [73, 74]. The CSP algorithm requires less computation load and is easy to implement. However, the CSP algorithm has several challenges. The first challenge is how to convert it to multiclass feature extraction. The second challenge is how to improve feature extraction accuracy.

(1) Multiclass CSP

One of the easiest methods to extend a feature extraction algorithm from binary class to multiclass is the "one versus rest" (OVR) technique. The OVA technique can be considered as an alternative form of a binary class algorithm because it considered one class as positive and all the other classes as negative. The OVR-based CSP algorithm was mentioned by Dornhege et al., and the mathematical derivation was explained by W. Wu et al. [75, 76]. The other method for extending a conventional CSP algorithm to multiclass is joint approximate diagonalization (JAD) [77]. The idea of the JAD-based CSP algorithm is to approximately diagonalize multiple covariance matrices for CSP transformation. The JAD-based CSP algorithm is widely used in driver studies because it improves the extraction accuracy compared with the OVR technique and has a relatively low computation load.

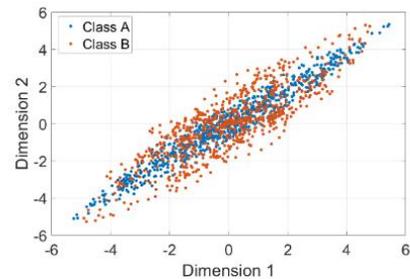

Fig 15a. 2 Classes EEG Signals Before CSP

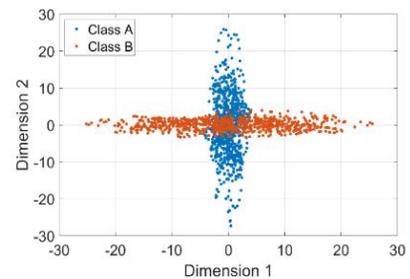

Fig 15b. Two Classes EEG Signals After CSP

(2) Extraction Accuracy Improvement

The CSP algorithm extraction accuracy can be improved by employing the filter bank technique, adaptive estimation, and nonparametric analysis. The filter bank-based CSP (FBCSP) was first proposed by K. K. Ang et al. in 2008 for motor imagery [78]. Then, this algorithm was applied for driver cognitive load analysis. The FBCSP algorithm analyzes and extracts features from the EEG signals through different frequency bandwidths. In addition, every extracted feature from the frequency band is classified through a classifier. Comprehensive scanning and classification ensure that the FBCSP algorithm picks the best features. However, this process also increases the algorithm processing time, especially when the classifier is a nonlinear classifier. The adaptive-based CSP algorithm (ACSP) was proposed by X. Song [79]. This algorithm combines the adaptive parameter estimation technique with the CSP algorithm to achieve better feature extraction performance. A. Costa et al. also designed an ACSP feature extraction algorithm [80]. They found that the ACSP algorithm is able to achieve similar classification results with fewer calibration sessions. The nonparametric analysis of CSP is based on the assumption of the non-Gaussian distribution of EEG data during driving [81]. According to the experimental results, the nonparametric analysis based CSP classification performance is 5% higher than that of conventional CSP.

With the abovementioned modification algorithms, the CSP feature extraction technique can extract driver state features effectively. However, the only constraint of all CSP algorithms is that the collected data must be multichannel because the CSP is a spatial filter-based feature extraction algorithm.

*b)    Spectral Analysis*

As shown in Table IV, the spectral analysis contains fast Fourier transform (FFT), power spectral density (PSD) analysis, and time-frequency analysis.

(1)    Fast Fourier Transform (FFT) Analysis

Fourier transform is one of the most common techniques to inspect signals in the frequency domain [82], as shown in Equations (8) and (9).

$$F\{g(t)\} = G(f) = \int_{-\infty}^{\infty} g(t) * e^{i*-2\pi ft} dt \quad (8)$$

$$F^{-1}\{G(f)\} = \int_{-\infty}^{\infty} G(f) * e^{i*-2\pi ft} df = g(t) \quad (9)$$

where $G(f)$ is the signal in the frequency domain and $g(t)$ is the signal in the time domain. FFT is an efficient algorithm for processing the Fourier transform in discrete time and discrete frequency (sampled frequency). When dealing with EEG data, the FFT algorithm transforms the time-series signal into the frequency domain, and the mean powers from different rhythms are selected as features. B. Peters et al. conducted FFT analysis and used the dominant frequency, average power, and center of gravity of frequency as features to detect the driver fatigue level [47]. C. Lin studied the mean power from FFT analysis to detect driver distraction [83].

The benefit of the FFT algorithm is its fast processing speed and ease of use. For instance, MATLAB has the built-in function "fft()" for fast Fourier transform, and in Python, the NumPy library also provides the "fft" function [56, 84].

The major disadvantage of the FFT algorithm is the loss of time-domain information. It is known that EEG signals are nonstationary [85]. Hence, the FFT transform cannot provide users with both temporal and frequency domain information for optimal spectral selection.

(2)    Power Spectral Density (PSD)

The PSD algorithm measures the power density of the EEG signal over a certain frequency band selected by the user. For driver state studies, frequency bands ranging from 8-30 Hz (alpha and beta rhythms) are popular choices. The maximum value, variance, and mean are usually selected as features [86]. Fig 16 presents PSD results from testing subject EEG data. According to Fig 16, the power density increases between 5-15 Hz (alpha and partial beta rhythm). PSD can be calculated through the FFT, Welch, and Burg methods. S. Yaacob compared different PSD methods for driver behavior studies [87]. According to their experimental results, the PSD analysis with the Welch method performs better in the detection of driver fatigue.

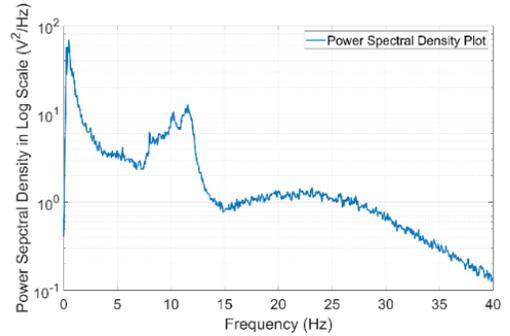

Fig 16.  EEG PSD analysis result

The PSD algorithm analysis procedures are simple and ready for real-time processing, which makes it one of the most common EEG-based driver state feature extraction techniques [88]. In addition, numerous studies illustrate how to tune parameters such as the time window and overlap percentage to improve extraction efficiency [89-91]. However, conventional PSD methods are not suitable for short data segments and sharp variations in the spectra [92, 93]. For driver state analysis, data segments are short, and sharp changes in spectra exist. Therefore, the autoregressive model-based PSD analysis is introduced.

The autoregressive model (AR) is another approach for PSD analysis. The AR model is a linear regression-based method for future signal estimation based on the present and previous signals, as shown in Equation (10):

$$X_t = \sum_{i=1}^{n} \varphi_i X_{t-i} + \varepsilon_t \quad (10)$$

where $\varphi_i$ is the corresponding AR parameter, $X_{t-i}$ is the current and previous observations, and $\varepsilon_t$ represents signal noise. The benefit of the AR model-based PSD analysis is its computational efficiency. H. Nguyen et al. applied the AR model for driver fatigue feature extraction [94]. According to their results, the classification accuracy was above 90%.

(3) Time-Frequency Analysis (TF Analysis)

TF analysis is a spectral analysis method that presents the signal spectral power in both the time and frequency domains. Since EEG signals during vehicle driving are nonstationary, the TF analysis could provide comprehensive temporal and frequency domain information simultaneously about the driver's state. The two most popular TF analysis methods are the short-time Fourier transform (STFT) and wavelet transform (WT). The STFT conducts a Fourier transform by selecting small time windows and composing all time windows together. The equation of the continuous-time STFT is

$$\{x(t)\}(\tau, \omega) = \int_{-\infty}^{\infty} x(t)\omega(t-\tau)e^{-j\omega t}dt \quad (11)$$

where $x(t)$ is the time-domain EEG signal and $\omega(t-\tau)$ is the time window function. For discrete-time STFT, the equation is

$$\{x[n]\}(k, \omega) = \sum_{n=-\infty}^{\infty} x[n]\omega[n-k]e^{-j\omega n} \quad (12)$$

where $k$ is the time resolution and the other parameters are the same as in Equation (11). The theory of the STFT algorithm is straightforward, and the implementation is simple. Thus, the STFT algorithm is widely applied in EEG-based driver state studies [95-98]. However, in STFT analysis, there is a trade-off between the time and frequency resolution. Both time and frequency resolutions are controlled by the selected time window. A large time window width provides a smaller frequency resolution but a larger time resolution and vice versa. Hence, some EEG information can be ignored with an improper choice of time window width. In addition, the time window width remains constant during the entire processing period, which is a waste of time when the STFT processes a nonimportant frequency band or time band. The WT algorithm can overcome the abovementioned drawbacks. As mentioned in the preprocessing chapter, WT includes CWT and DWT processing. In the context of EEG-based driver state studies, DWT is commonly used for feature extraction. The DWT equation and processing procedures are shown in Equation (5) and Fig 8, respectively. The determination of the DWT decomposition level is trivial, and the corresponding literature is tabulated in Table IV [99-105].

*2) Statistical-based Methods*

Unlike signal-processing-based methods, statistical-based methods do not study signal dynamics but detect and extract EEG driver state-related features from a large amount of experimental data. Usually, the use of statistical analysis methods does not require too much prior knowledge about EEG features and driver-related EEG behaviors. However, without applying any prior knowledge about the signal may encounter a high computational load issue. In this section, discriminant analysis and statistical entropy analysis methods are illustrated.

*a) Discriminant Analysis*

Discriminant analysis is a statistical technique to project samples with different classes to a hyperplane so that the distance between the data in each class and the hyperplane becomes maximum. Fig 17 illustrates the discriminant analysis principal in the 3D animation.

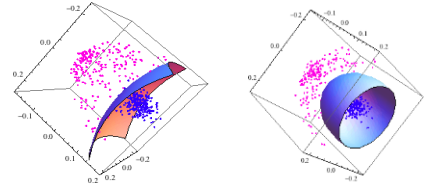

Fig 17. Discriminant Analysis Hyperplane and Data Presentation

Discriminant analysis is commonly applied for driver fatigue feature extraction. C. Lin employed a nonparametric feature extraction method for motion sickness detection study [106]. He also compared the nonparametric discriminant analysis method with the linear discriminant method and found that the nonparametric discriminant analysis exhibits 20% better classification performance than the LDA method [107]. According to the literature results, discriminant analysis provides excellent feature extraction performance.

*b) Statistical Entropy*

Entropy, originally, was a measure of the disorder of a system in the thermodynamics field. In 1948, C. Shannon introduced the entropy theory into the information and signal processing field, which was named statistical entropy [108]. The equation is

$$H = -\sum_{i=1}^{n} P_i \log(P_i) \quad (13)$$

where $P_i$ is the probability of case $i$ in the event. For EEG application, the entropy method is used for quantification of the similarity among every selected EEG pattern in either the time or frequency domain.

The most popular algorithms for statistical entropy analysis in the time domain are approximate entropy and sample entropy. The approximate entropy (AE) equation is

$$AE(m, r) = \frac{1}{N-m+1}\Phi^m - \frac{1}{N-m+2}\Phi^{m+1} \quad (14)$$

where $N$ is the length of the EEG data, $m$ is the selected EEG segment data length, and $\Phi^m$ and $\Phi^{m+1}$ are the average of similarity fractions as shown below:

$$\Phi^m = \sum_{i=1}^{N-m+1} \log(C(i,r,m)) \quad (15)$$

$$\Phi^{m+1} = \sum_{i=1}^{N-m} \log(C(i,r,m)) \quad (16)$$

where $C(i,r,m)$ is the similarity between the selected EEG segments and the $m^{th}$ segments. The similarity threshold is a user predefined value $r$. The smaller the AE value, the higher the repeatability in the signal. The advantages of approximate entropy analysis are the tolerance of noisy signals and nonprior knowledge requirements about the signals [109]. However, since approximate entropy requires a self-check analysis (the selected EEG segments have to be compared with their own so that special cases that cause log (0) do not exist), the result is biased, which is a critical issue when dealing with a small number of EEG segments. The sample entropy (SE) is introduced to solve the potential biased results issue, and the equation is

$$SE(m,r) = -\log(\Phi^m/\Phi^{m+1}) \quad (17)$$

where $\Phi^m$ and $\Phi^{m+1}$ are the same as Equations (15) and (16). Similar to AE, the smaller the SE value, the higher the repeatability of the signal.

In the frequency domain, spectral-based entropy (SPE) and wavelet-based entropy (WE) are commonly applied [110, 111]. SPE can be considered an extension of PSD analysis, and the equation is

$$SPE = -\sum_{i=1}^{n} P(i) \log(P(i)) \quad (18)$$

where $P(i)$ is the normalized power spectral density, as shown in Equation (19):

$$P(i) = \frac{A(\omega_i)}{\sum_i A(\omega_i)} \quad (19)$$

where $A(\omega_i)$ is the PSD at the $i^{th}$ frequency range and $\sum_i A(\omega_i)$ is the sum of PSD among all frequency ranges. The WE analysis requires wavelet transform. After conducting the wavelet transform, the EEG energy in the user-defined decomposed level can be calculated as

$$P(j) = \frac{E(j)}{\sum E(j)} \quad (20)$$

where $E(j)$ is the energy at level $j$ and $\sum E(j)$ is the sum of energy among all levels. Then, the WE is calculated as

$$WE = -\sum_{i=1}^{n} P_i \log(P_i) \quad (21)$$

Both SPE and WE can extract EEG features effectively, but it requires frequency or time-frequency analysis and prior knowledge about the signals.

Currently, the applications of statistical entropy on EEG analysis are combinations with AE, SE, SPE, and WE. P. Wang applied SPE, AE, SE, and fuzzy entropy to detect driver fatigue [112]. According to his findings, the combination of all entropy feature extraction methods could provide better detection accuracy. In addition to combination features, modification and improvement based on current entropy analysis are also popular. H. Wang combined WE, AE, and SE as features for driver fatigue studies [104]. In the research, they introduced the term "peak-to-peak entropy" to further improve the extraction effectiveness, as shown below:

$$PP_{en} = \max(En(t)) - \min(En(t)) \quad (22)$$

where $En(t)$ is either AE or SE. A. Routray proposed chaotic entropy for driver drowsiness detection [113]. The chaotic entropy analysis is composed of AE, SE, and a modified SE algorithm (MSE), whose equation is

$$mSE(m,r) = \frac{1}{\left(1 + e^{\frac{d[x(i),x(j)]-0.5}{r}}\right)} \quad (23)$$

where $d[x(i), x(j)]$ is the distance between the EEG segments and $r$ is the tolerance threshold. The objective of the chaotic entropy analysis is to select the most effective feature from AE, SE, and MSE for different EEG electrodes to improve the classification accuracy.

TABLE IV
EEG COMMON FEATURE AND POPULAR FEATURE EXTRACTION ALGORITHM SUMMARY

| EEG Common Features | |
|---|---|
| **Feature Name** | **Feature Observation** |
| ERD/ERS, ERP, mean, variance | EEG signals fluctuation in temporal domain due to events stimulus |
| Frontal, Occipital, and Parietal Lobe | Key brain regions for driver vigilance, drowsiness, or distraction detection |
| Beta Rhythm, Alpha Rhythm | Key brain rhythms for driver vigilance, drowsiness, or distraction detection |

| EEG Feature Extraction Techniques | | | |
|---|---|---|---|
| **Algorithm Name** | | **Analysis of Substance** | **Related Research** |
| Signal Processing based Methods: Analyze based on signal processing technique from time, frequency, or spatial domain | | | |
| | Common Spatial Pattern | Spatial filter finding the maximum difference among classes | [72-81] |
| | Power Spectral Density | Signal power analysis over certain frequency band | [86-93] |
| Spectral Analysis | Fast Fourier Transform | Calculate signal power or energy over pre-defined time segmentation | [47], [56], [82-84] |
| | Time-Frequency Analysis | Short-time Fourier transform for non-stationary EEG signals | [95-98] |
| | | Wavelet transform for non-stationary EEG signals | [99-105] |
| Statistical-based Methods: Pattern Recognition based on Statistical Analysis | | | |
| | Discriminant Analysis | Bi/multi-class EEG signals projected to a hyperplane for maximum separation | [106][107] |
| | Statistical Entropy | Find correlations among all EEG signals based on user defined time/frequency segment | [104][109-112] |

## V. EEG CLASSIFICATION METHODS

The classification algorithm of EEG analysis in driver condition studies can be categorized as a traditional machine learning classifier and deep learning-based classifier. The traditional methods are mostly state-of-art classifiers such as linear discriminant analysis (LDA), support vector machine (SVM), naïve Bayes (NVB), and k$^{th}$ nearest neighbor (kNN). The deep learning methods are different neural network models such as the feedforward neural network (FFN), convolutional neural network (CNN), recurrent neural network (RNN), fuzzy neural network (FNN), and autoencoder neural network (AE). Table V presents a summary and brief explanation of every method, and their detailed explanation and equations are illustrated below.

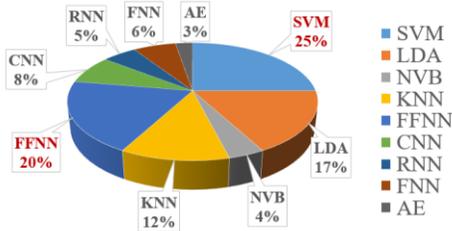

Fig 18. EEG-based Driver State Analysis Classification Algorithm Distribution

### A. Traditional Machine Learning Algorithms

#### 1) Linear Discriminant Analysis

The LDA algorithm belongs to discriminant analysis, which is one of the simplest but most effective classifiers in real-world applications. It linearly transforms data to a hyperplane to maximize the distance among each class of data. The LDA algorithm finds the transformation matrix by solving and finding the maximum eigenvalue from the between-class and with-in class metrics. The equation for the maximum eigenvalue is

$$E = \max\left(eig(S_w^{-1} S_B)\right) \quad (24)$$

where $S_w$ and $S_B$ are the with-in and between-class metrics, respectively. Their equations are

$$S_w = \sum_{i=1}^{n}(x_i - \mu_{yi})(x_i - \mu_{yi})^T \quad (25)$$

$$S_b = \sum_{k=1}^{m}(\mu_k - \mu)(\mu_k - \mu)^T \quad (26)$$

where $x_i$ is the collection of the $i^{th}$ sample, $\mu_{yi}$ is the mean value of the $i^{th}$ sample, and $\mu$ is the overall mean value of all samples. After finding the maximum eigenvalue, the corresponding eigenvectors compose the transformation matrix.

The desired classifier for EEG-based driver state study requires i) ease of implementation and ii) robustness to different types of features. The LDA classifier meets both requirements. For ease of implementation, there are numerous built-in and open-source LDA classifier packages, such as the "scikit-learn" library for Python and the "fitcdiscr" function in MATLAB. For the robustness of different features, the LDA algorithm can maintain high classification accuracy for almost all different types of input EEG features. J. Milan, et al. employ the "Cz" electrode potentials as the features for LDA classification for driver behavior studies [114]; K. K. Ang, et al. used CSP as the feature for LDA classification in driver cognitive load research [115]; X. Fan, et al. applied theta wave power as input for LDA classification to recognize emergency situations [116]. The abovementioned literature selects different features for the LDA classifier but still obtains a classification accuracy over 70%.

However, the LDA classifier can only predict discrete classes such as drowsiness and non-drowsiness or level 1 distraction, level 2 distraction, and alert. Hence, the LDA algorithm is not fit for continuous regression analysis. Moreover, the LDA is sensitive to outlier data. If the extracted EEG features are noisy, the LDA prediction performance will decrease. Therefore, a careful artifact removal and signal preprocessing process is necessary for LDA classification.

## 2) Support Vector Machine (SVM)

SVM, similar to LDA, defines a hyperplane to separate different classes of data at a maximum distance. However, the main differences between SVM and LDA are as follows: first, the SVM only considers the data points near the classification boundaries; second, SVM can be tuned by a kernel function to dramatically improve its classification accuracy. A.K. Nagar proposed using kernel-based SVM to detect driver cognitive state [105]. They found that the SVM with a radial-based function (RBF) kernel classification accuracy is 2% higher than the SVM with a linear kernel function. J. Zhang proposed an RBF kernel-based support vector machine combined with a particle swarm optimization (PSO) algorithm to increase the driver vigilance prediction accuracy [95]. In addition to the kernel function, J. Lian proposed a transductive support vector machine with prior information (PI-TSVM) to improve the classification accuracy [117]. The TSVM is similar to a semi-supervised learning algorithm but learning is based on the test data. They collected the ratio of the positive samples to negative samples as prior information. With the help of this information and its application to the TSVM, the classification accuracy is improved.

The major benefit of the SVM classifier is the classification accuracy, especially with the help of kernel functions. According to J. Lin and J. Zhang's experimental results, the SVM with kernel function classification accuracy is similar to the deep learning algorithm, and the training period is shorter than the deep learning algorithm. Moreover, there are many existing open-source SVM algorithms available, such as "fitcsvm" for SVM and LIBSVM for C++ and Java [118]. Hence, implementing the SVM classifier is easy and effective. However, when tuning the SVM parameters, the selection of a kernel function is time consuming and trivial. Therefore, a proper SVM classifier for EEG-based driver state study requires experience and prior knowledge about the EEG signals.

## 3) Naïve Bayes (NVB)

NVB is a probabilistic-based classifier for multidimensional data classification. The equation is

$$P(X|Y) = \frac{P(Y|X)P(X)}{P(Y)} \quad (27)$$

where $X$ is the target event and $Y$ is the known event. The assumption of the NVB classifier is that every input feature is independent of each other and that the contributions among every feature are equal. Despite easy application and multivariate analysis, the NVB method did not present good classification results for most EEG-based driver state studies. C. Lin applied the NVB classifier to study driver cognitive state [106]. According to the experimental results, the NVB method-based classification results are lower than those of the other state-of-art methods. J. Hu proposed an algorithm for automatic detection of driver fatigue [119]. They compared the NVB classified results with the AdaBoost and SVM classifiers through the receiver operating characteristic (ROC) curve method. His results have demonstrated that both SVM and AdaBoost classification results are over 30% better than the NVB classifier. The main reasons for the poor NVB performance are the independent feature input and the equal feature weight requirements. In reality, the collected EEG signal in every channel is a combination of multiple independent components. Hence, using the EEG time-domain signal as an input for NVB classification violates the independent rule. Even though ICA preprocessing can be conducted, the ground truth about the number of independent components is unknown. Therefore, NVB still cannot perform an ideal classification. Moreover, the features extracted weight from EEG cannot be guaranteed to be equal because the features exhibit different characteristics with different testing subjects or testing environments. Therefore, despite its ease of application and fewer training period advantages, the NVB classifier usually cannot provide the best classification results in EEG-based driver state studies.

## 4) $K^{th}$ Nearest Neighbor (KNN)

The kNN is a nonparametric classification algorithm based on distance estimation and majority vote. The basic idea about the kNN is to calculate the distances between the unknown data and known data and then rank the distances from low to high. The results of the unknown data class are determined by the highest vote in the first $k^{th}$ close distance known data. There are three common distance functions: Euclidean, Manhattan, and Minkowski, which are presented below [120].

$$D_{Eu} = \sqrt{\sum_{i=1}^{k}(x_i - y_i)^2} \quad (28)$$

$$D_{Ma} = \sum_{i=1}^{k}|x_i - y_i| \quad (29)$$

$$D_{Mi} = \left(\sum_{i=1}^{k}(|x_i - y_i|)^q\right)^{\frac{1}{q}} \quad (30)$$

where $x$ is the testing sample, and $y_i$ is the known sample.

The main benefit of the kNN algorithm is no training period. There is no training phase in the kNN method but directly calculates the distances between the testing data and the existing known data. Moreover, the kNN algorithm could be considered a local optimization algorithm. In EEG signals, some features are locally clustered. Therefore, this method is suitable. However, the main disadvantage of kNN is that it is time consuming when dealing with a multidimensional large dataset. Since every point distance is required for calculation, the computation time is dramatically increased. The other disadvantage of the kNN algorithm is the biased distance when dealing with abnormal features. The solution to this is standardization. S. H. Adil compared the kNN, SVM, and NVB classifiers to detect driver distraction [121].

## B. Deep Learning Algorithms

Originally, deep learning-based classification algorithms were applied in image processing. Recently, it has become a popular classification technique in EEG driver state detection with the development of computation ability. The deep learning algorithms estimate parameters from training data through massive computation. Therefore, generally speaking, the deep learning-based algorithm classification results are accurate. In this section, the feedforward neural network (FNN), convolutional neural network (CNN), recurrent neural network (RNN), and autoencoder (AE) are illustrated.

### 1) Feedforward Neural Network (FFNN)

The FFNN is the most basic but effective neural network topology for EEG-based driver state analysis, as shown in Fig 19. The FFNN combines the input directly with weight and bias factors to calculate the output with the equation shown below:

$$y_j = \sum_{i=1}^{n} w_{ji} x_i + b_i \qquad (31)$$

where $w_{ji}$ is the weight factor in the $j^{th}$ layer and $i^{th}$ neuron and $b_i$ is the bias factor for the $i^{th}$ neuron. To obtain classification results, an activation function such as sigmoid or ReLU is used after calculating the output ($y_i$) [122, 123]. For EEG-based driver state analysis, the FFNN inputs are either EEG signals or features extracted from the raw EEG signals, and the outputs are desired classification results such as fatigue or distraction level. The advantage of the FFNN is the robustness for different types of feature input. However, the FFNN model requires massive computation power to estimate the weight and bias factor. Moreover, the selection of neurons and layers can cause overfitting issues. Just like any other learning method, the FFNN requires large training sets to learn the content of the input signals in order to produce (fire) an output desirably. The next paragraphs demonstrate the advantages and improvements of the FFNN model.

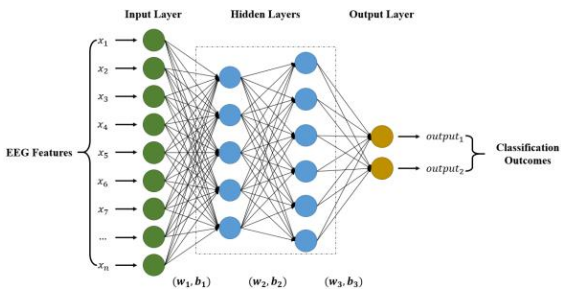
Fig 19. Typical FNN Model Topology

#### a)   Robustness to Different Types of Features

According to previous EEG driver state study literature, nearly all common EEG extracted features, such as time-series signals, DSP features, DWT features, and statistical entropy can be used for the FFNN model with relatively high classification accuracy. S. Yaacob employed PSD as a feature to classify driver fatigue [87]. A 4-layer FFNN topology with the error backpropagation learning algorithm was developed, and the classification rate achieved 85%. S. Mozaffari applied the signal mean, standard deviation, and power from DWT decomposition as features for the FFNN model to detect driver fatigue [102]. Their experimental results indicate that the accuracy is approximately 89%. In addition to the single feature input, hybrid combined features have also been selected. H. Wang et al. selected both sample entropy and wavelet entropy to feed into FNN for detecting driver drowsiness levels [124]. With the combination of entropy features, the classification accuracy is over 96.5%. The robustness of the FFNN model ensures that it becomes the most popular classifier in the EEG driver state study (Fig 18).

#### b)   Massive Computation Load and Overfitting

(1) Computation Load Reduction

The reason for the massive computational load is the backpropagation algorithm convergence rate. To solve this issue, E. Leber et al. applied Laverberg-Marquardt backpropagation (LMBP) [99]. The LMBP method, compared with the conventional error backpropagation (EBP) algorithm, has a faster convergence rate and is more robust. With this learning algorithm, the driver drowsiness detection rate is over 80%. Another learning optimization technique is the magnified gradient function (MGF). The objective of the MGF function is to speed up the convergence rate by magnifying the gradient function [125]. S. Lal et al. applied the MGF function for learning professional and unprofessional driver drowsiness states [126]. According to their results, both professional and unprofessional driver fatigue detection accuracy is higher than 80%.

(2) Overfitting Issue Prevention

Overfitting is a common issue for the FFNN model in EEG-based driver state studies because of the noisy input signals and sensitive classification. Bayesian FFNN (BFFNN) is a common neural network model to prevent overfitting. The BFFNN model predicts the weights and bias factor based on Bayesian probability, which regularizes the model. H. Nguyen applied the BFFNN model to classify driver fatigue. For their classification process, three layers' weight and bias factors are regularized based on the Bayes theorem. Their classification accuracy achieves close to 90%.

Even though the FFNN could be tolerated for multiple feature inputs, the detection results could be tuned by using different learning algorithms. The learning speed is slow and requires large testing data. Therefore, it cannot be used for real-time processing and classification.

*2) Convolutional Neural Network (CNN)*

The CNN model topology contains one or multiple convolutional layers and a fully connected neural network, as shown in Fig 20. In the convolutional layer, the kernel filter is convolved with the input data with the equation

$$G[m,n] = (f*h)[m,n] = \sum_j \sum_k h[j,k]f[m-j,n-k] \quad (32)$$

where $h$ is the kernel filter, and $f$ is the input signal. After the kernel filter, some CNN models down sample the extracted features for faster computational speed. In the fully connected neural network, the extracted features are applied for classification. Thus, in the context of EEG-based driver state study, the convolutional layers can replace feature extraction, and the fully connected network is considered as classification. The most common application of the CNN model for driver state study is the end-to-end learning technique, which means considering the raw EEG signal directly as input and the driver condition as output. W. Kameyam et al. employed CNN for end-to-end learning in driver workload studies [127]. In the CNN topology, eight convolutional layers were developed, and the outputs were "low", "medium", and "high" workload conditions. Their detection accuracy reached 93.4%. However, the traditional end-to-end CNN cannot extract implicit EEG features such as spatial features. Therefore, some studies have modified the CNN model to extract more implicit features to improve classification accuracy.

C. Lin et al. proposed a novel channel-wise CNN (CCNN) method for driver workload estimation [128]. They found that when predicting the driver cognitive load with raw EEG data, the CCNN detection accuracy outperforms the traditional machine learning algorithm and FFNN model. A similar modification of the CNN model was also applied by S. Zuo [129]. In their modified CNN topology, there were layers to extract EEG spatial and temporal features. The detection accuracy was compared with the SVM classifier, and the results were 2% higher. Y. Huang has applied covariance-based CNN, which achieved better results than other common CNNs [130].

Despite the high classification accuracy, the CNN topology requires a large quantity of data for training and a lengthy training period. Moreover, CNNs easily cause overfitting issues with a small amount of data [131]. Therefore, the CNN algorithm is not suitable for real-time unsupervised studies for EEG-based driver condition analysis.

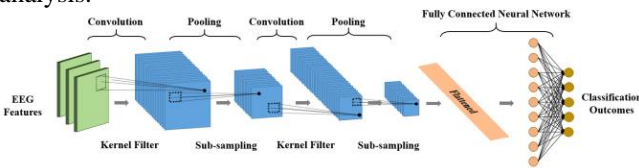

Fig 20. CNN Model Topology

*3) Recurrent Neural Network (RNN)*

The RNN model is developed based on the idea of parameter sharing [132]. The distinct part of the RNN topology is that neurons in the same layer are dependent on each other, as shown in Fig 21. Equations for RNN are shown below:

$$x(t) = \sigma(w^{in}u(t) + w^{rec}x(t-1) + b) \quad (33)$$

where $\sigma$ is the activation function, $w^{in}$ is the input weight factor for the previous layer, $w^{rec}$ is the recurrent weight factor for the previous perceptron, $u(t)$ is the input, $x(t-1)$ is the previous perceptron, and $b$ is the bias factor.

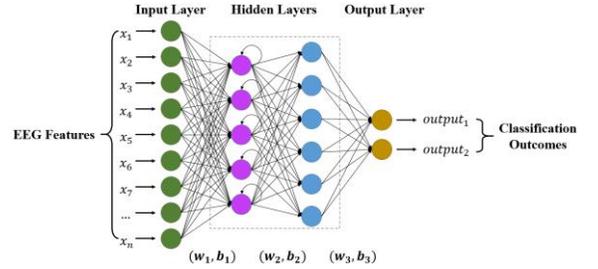

Fig 21. RNN Model Topology

Since the EEG electrode collected signals are a combination of multiple independent components, the EEG signals are dependent on each other. Therefore, the RNN topology can provide accurate detection accuracy. T. Falk et al. employed RNN for driver intention detection [133]. In this study, EEG signal features are extracted through a spike algorithm. The RNN model is composed of a reservoir and a readout layer where the reservoir contains temporal and spatial information. According to the experimental results, the detection accuracy reached over 88%, which is higher than that of other traditional machine learning techniques, such as SVM. The RNN model can also be combined with the CNN. In S. Lee's research on driver brake intention, they compared RCNN with the LDA classifier [134]. The area under the curve (AUC) for CNN was 0.86, which was 0.25 higher than that of the LDA classifier. The RNN model was also combined with a fuzzy neural network (FNN) model [135]. C. Lin proposed a recurrent self-evolving fuzzy neural network (RSEFNN) to increase the memory capability for noise cancellation. A special RNN, the Hopfield neural network, was employed by A. Nagar [136], and the equation [137] is

$$C\frac{d\vec{u}}{dt} = -\alpha\vec{u} + W\vec{x} + \vec{\theta} \quad (34)$$

where $\alpha$ is a diagonal factor matrix, $u$ is the system state vector, and $W$ is a symmetric weighted matrix. According to their experimental results, the Hopfield-based RNN model outperformed other learning models, such as KNN and SVM.

The disadvantages of the RNN model are the high computational load or unstable training due to the vanishing/exploding gradient issue. The nature of the RNN model is to contain previous time point neuron information for the next time point calculation. Thus, during the learning period, a close-to-zero random value multiplication causes a smaller gradient (vanishing gradient problem) and longer training time, while a large random value multiplication causes a larger gradient (exploding gradient problem) and the model is not stable. Therefore, the RNN model training period is time consuming compared with other neural network models.

*4) Fuzzy Neural Network (FNN)*

The FNN model combines the artificial neural network (ANN) and fuzzy systems, as shown in Fig 22. Both ANN and fuzzy systems are pattern recognition algorithms. The benefit of ANN is that it does not require prior knowledge. However, ANN requires a large number of observations, and the training process is not straightforward. Fuzzy systems do not require a large training dataset, and the learning process is clear. Nevertheless, fuzzy systems require prior knowledge about the learning data, and the training can be time consuming. Hence, a combination of ANN and FNN could unite both advantages but exclude the disadvantages.

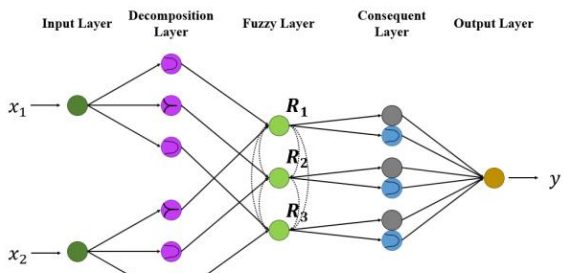

Fig 22. FNN Model Topology

Lin et al. proposed a novel channel-wise CNN (CCNN) method for driver workload estimation [128]. They found that when predicting the driver cognitive load with raw EEG data, the CCNN detection accuracy outperforms the traditional machine learning algorithm and FFNN model. A similar modification of the CNN model was also applied by S. Zuo [129]. In their modified CNN topology, there were layers to extract EEG spatial and temporal features. The detection accuracy was compared with the SVM classifier, and the results were 2% higher. Y. Huang has applied covariance-based CNN, which achieved better results than other common CNNs [130].

*5) Autoencoder (AE)*

AE is an unsupervised deep learning algorithm that estimates the output so that it is similar to the input. The topology of the AE algorithm is shown in Fig 23.

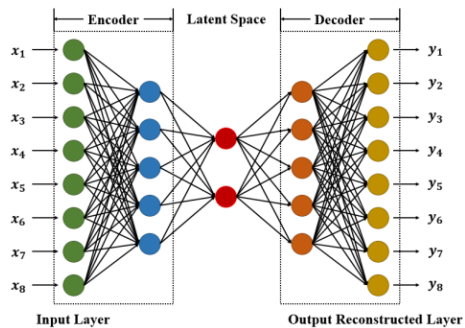

Fig 23. Autoencoder Network Model Topology

J. Lian et al. applied variational AE to provide a robust feature representation of EEG signals [117]. In their study, the AE was combined with transductive SVM (TSVM). The semi-supervised learning algorithm only requires small initial training data and can continue training with unlabeled EEG signals. Presently, the autoencoder classifier is not widely applied for EEG-based driver state studies. However, because of the unsupervised learning characteristics, the autoencoder classifier exhibits a promising future in EEG analysis.

TABLE V
CLASSIFICATION ALGORITHM SUMMARY TABLE

| Algorithm Name | Analysis of Substance | Related Literatures |
|---|---|---|
| **Traditional Machine Learning Algorithms** | | |
| Linear Discriminant Analysis | Linearly transform input data to maximize the distance among different classes | [113-115] |
| Support Vector Machine | Estimate a hyperplane to maximize the margin distance among different classes | [116-118] |
| Naïve Bayes | Statistical estimation for multi-dimensional data | [119] |
| Kth Nearest Neighbor | Majority vote based classification rule without training phase | [120-121] |
| **Deep Learning Algorithms** | | |
| Feedforward Neural Network | Most basic neural network model by predicting input data classes with estimated weight and bias factors | [123-126] |
| Convolutional Neural Network | Convolving input and estimated kernel filters to achieve both feature extraction and classification | [127-130] |
| Recurrent Neural Network | Estimating weight and bias factor from both the input and previous neurons in the same layer | [131-134] |
| Fuzzy Neural Network | Neural network model combined with fuzzy rule | [135-140] |
| Auto Encoder | Unsupervised learning algorithm for data reconstruction so that the output is similar to the input | [141] |

## VI. CONCLUSIONS AND FUTURE WORKS

In this paper, the EEG systems for driver state analysis and the corresponding EEG state-of-art algorithms are reviewed. According to previous literature, EEG driver state analysis systems tend toward convenient wearing, compact carrying, and economic pricing. However, the reliability, external noise resistance, and wireless connection of EEG driver state analysis systems still need to be improved. The EEG-based brain wave signal monitoring approaches are reviewed with sufficient depth to familiarize a reader with the available tools and their advantages and limitations. The methods for signal collections, signal pre-filtering, signal processing, signal feature extractions and classifications are reviewed. The previous studies with some successes in each subarea for driver state monitoring are presented, along with listing of any limitations where applicable. The development of EEG driver state analysis algorithms for comprehensive driver state detection, accurate classification accuracy, and efficient computational load are discussed. Although substantial progress in EEG driver state analysis has been made, EEG signal analysis algorithms still need to be enhanced in three fields: EEG artifact reduction, real-time processing, and between-subject classification accuracy. For EEG artifact reduction, although numerous artifact removal algorithms exist, these algorithms either cannot remove all EEG artifacts thoroughly or overfitting useful EEG information. For real-time processing, most published algorithms achieve real-time processing speed by sacrificing the algorithm effectiveness. For the between-subject classification accuracy issue, almost all existing EEG-based driver state algorithms cannot be implemented for subject-independent analysis.

In brief, by solving the challenges of EEG systems and corresponding driver state analysis algorithms, a deeper understanding of driver state and control functions and readiness can be developed. This can assist in developing ADAS or semi-autonomous systems to minimize human error during driving and enhance safety.

## REFERENCES

[1] NHTSA, "2016 Fatal Motor Vehicle Crashes: Overview," 2017.
[2] A. Broggi *et al.*, "PROUD-Public road urban driverless test: Architecture and results," in *2014 IEEE Intelligent Vehicles Symposium Proceedings*, 8-11 June 2014 2014, pp. 648-654, doi: 10.1109/IVS.2014.6856478.
[3] V. D. Pyrialakou, C. Gkartzonikas, J. D. Gatlin, and K. Gkritza, "Perceptions of safety on a shared road: Driving, cycling, or walking near an autonomous vehicle," *Journal of Safety Research,* vol. 72, pp. 249-258, 2020/02/01/ 2020, doi: https://doi.org/10.1016/j.jsr.2019.12.017.
[4] M. Q. Khan and S. Lee, "A Comprehensive Survey of Driving Monitoring and Assistance Systems," (in eng), *Sensors (Basel),* vol. 19, no. 11, p. 2574, 2019, doi: 10.3390/s19112574.
[5] "LS 600h L Owner's Manual," 2008.
[6] BMW, "Owner's Manual The BMW X5," 2019.
[7] "Tesla Model 3 Owner's Manual," 2019.
[8] S. N. Abdulkader, A. Atia, and M.-S. M. Mostafa, "Brain computer interfacing: Applications and challenges," *Egyptian Informatics Journal,* vol. 16, no. 2, pp. 213-230, 2015/07/01/ 2015, doi: https://doi.org/10.1016/j.eij.2015.06.002.
[9] S. F. Liang, C. T. Lin, R. C. Wu, Y. C. Chen, T. Y. Huang, and T. P. Jung, "Monitoring Driver's Alertness Based on the Driving Performance Estimation and the EEG Power Spectrum Analysis," in *2005 IEEE Engineering in Medicine and Biology 27th Annual Conference*, 17-18 Jan. 2006 2005, pp. 5738-5741, doi: 10.1109/IEMBS.2005.1615791.
[10] F. Wang, Q. Xu, and R. Fu, "Study on the Effect of Man-Machine Response Mode to Relieve Driving Fatigue Based on EEG and EOG," (in eng), *Sensors (Basel),* vol. 19, no. 22, p. 4883, 2019, doi: 10.3390/s19224883.
[11] M. Lemke, "Correlation between eeg and driver's actions during prolonged driving under monotonous conditions," *Accident Analysis & Prevention,* vol. 14, no. 1, pp. 7-17, 1982/02/01/ 1982, doi: https://doi.org/10.1016/0001-4575(82)90003-3.
[12] K. Idogawa, S. P. Ninomija, and F. Yano, "A time variation of professional driver's EEG in monotonous work," in *Images of the Twenty-First Century. Proceedings of the Annual International Engineering in Medicine and Biology Society*, 9-12 Nov. 1989 1989, pp. 719-720 vol.2, doi: 10.1109/IEMBS.1989.95949.
[13] S. K. L. Lal, A. Craig, P. Boord, L. Kirkup, and H. Nguyen, "Development of an algorithm for an EEG-based driver fatigue countermeasure," *Journal of Safety Research,* vol. 34, no. 3, pp. 321-328, 2003/08/01/ 2003, doi: https://doi.org/10.1016/S0022-4375(03)00027-6.
[14] D. Göhring, D. Latotzky, M. Wang, and R. Rojas, "Semi-autonomous Car Control Using Brain Computer Interfaces," in *Intelligent Autonomous Systems 12*, Berlin, Heidelberg, S. Lee, H. Cho, K.-J. Yoon, and J. Lee, Eds., 2013// 2013: Springer Berlin Heidelberg, pp. 393-408.
[15] A. Eskandarian, "Safety issues of Drowsy/Fatigue Driving and Countermeasure mitigation," presented at the Road safety on four continents: 15th international conference, Abu Dhabi, United Arab, 2010.
[16] R. A. Sayed, A. Eskandarian, and P. Delaigue, "Driver fatigue: causes and countermeasures," 2005.
[17] J. D. M. Lee, Jane; Brown, Timothy L; Roberts, Shannon C; Schwarz, Chris; Yekhshatyan, Lora; Nadler, Eric; Liang, Yulan; Victor, Trent; Marshall, Dawn; Davis, Claire, "Distraction Detection and Mitigation Through Driver Feedback," 2013.
[18] A. Eskandarian, R. A. Sayed, P. Delaigue, A. Mortazavi, and J. J. Blum, "Advanced Driver Fatigue Research," 2007.
[19] A. Eskandarian and A. Mortazavi, "Evaluation of a Smart Algorithm for Commercial Vehicle Driver Drowsiness Detection," *2007 IEEE Intelligent Vehicles Symposium,* pp. 553-559, 2007.
[20] A. Eskandarian, *Handbook of Intelligent Vehicles*, 1 ed. London: Springer-Verlag London, 2012, p. 1599.
[21] A. Eskandarian, Ali Mortazavi, "Unobtrusive Driver Drowsiness Detection System and Method," United States, 2009.
[22] M. Guo, S. Li, L. Wang, M. Chai, F. Chen, and Y. Wei, "Research on the Relationship between Reaction Ability and Mental State for Online Assessment of Driving Fatigue," (in eng), *Int J Environ Res Public Health,* vol. 13, no. 12, p. 1174, 2016, doi: 10.3390/ijerph13121174.
[23] D. McCarthy, "Taxonomy of Older Driver Behaviors and Crash Risk," 2018.
[24] J. G. Gaspar and D. V. McGehee, "Driver brake response to sudden unintended acceleration while parking," *Transportation Research Interdisciplinary Perspectives,* vol. 2, p. 100039, 2019/09/01/ 2019, doi: https://doi.org/10.1016/j.trip.2019.100039.
[25] Y. Xing, C. Lv, and D. Cao, "Chapter 5 - Driver Behavior Recognition in Driver Intention Inference Systems," in *Advanced Driver Intention Inference*, Y. Xing, C. Lv, and D. Cao Eds.: Elsevier, 2020, pp. 99-134.
[26] J. Kim, H. Suk, J. Kim, and S. Lee, "Combined regression and classification approach for prediction of driver's braking intention," in *The 3rd International Winter Conference on Brain-Computer Interface*, 12-14 Jan. 2015 2015, pp. 1-3, doi: 10.1109/IWW-BCI.2015.7073027.
[27] I.-H. Kim, J.-W. Kim, S. Haufe, and S.-W. Lee, "Detection of braking intention in diverse situations during simulated driving based on EEG feature combination," *Journal of Neural Engineering,* vol. 12, no. 1, p. 016001, 2014/11/26 2014, doi: 10.1088/1741-2560/12/1/016001.
[28] T. Teng, L. Bi, and Y. Liu, "EEG-Based Detection of Driver Emergency Braking Intention for Brain-Controlled Vehicles," *IEEE*

*Transactions on Intelligent Transportation Systems,* vol. 19, no. 6, pp. 1766-1773, 2018, doi: 10.1109/TITS.2017.2740427.

[29] K. J. Haufe S, Kim IH, Sonnleitner A, Schrauf M, Curio G, Blankertz B, "EEG-based detectionof emergency braking intention in real-world driving," *Journal of Neural Engineering,* vol. 11, no. 5, 2014.

[30] P. Riyahi, "A Steady-State Visual Evoked Potential Brain-Computer Interface System Evaluation as an In-Vehicle Warning Device," Ph.D. Thesis, 2014. [Online]. Available: https://ui.adsabs.harvard.edu/abs/2014PhDT.......156R

[31] "Tricks and Tips in BrainVision Recorder." https://pressrelease.brainproducts.com/recorder-tips-2/ (accessed.

[32] A. Delorme and S. Makeig, "EEGLAB: an open source toolbox for analysis of single-trial EEG dynamics including independent component analysis," *Journal of Neuroscience Methods,* vol. 134, no. 1, pp. 9-21, 2004/03/15/ 2004, doi: https://doi.org/10.1016/j.jneumeth.2003.10.009.

[33] V. Peterson, C. Galván, H. Hernández, and R. Spies, "A feasibility study of a complete low-cost consumer-grade brain-computer interface system," *Heliyon,* vol. 6, no. 3, p. e03425, 2020/03/01/ 2020, doi: https://doi.org/10.1016/j.heliyon.2020.e03425.

[34] K. E. Mathewson, T. J. L. Harrison, and S. A. D. Kizuk, "High and dry? Comparing active dry EEG electrodes to active and passive wet electrodes," *Psychophysiology,* vol. 54, no. 1, pp. 74-82, 2017/01/01 2017, doi: 10.1111/psyp.12536.

[35] E.-R. Symeonidou, A. D. Nordin, W. D. Hairston, and D. P. Ferris, "Effects of Cable Sway, Electrode Surface Area, and Electrode Mass on Electroencephalography Signal Quality during Motion," (in eng), *Sensors (Basel),* vol. 18, no. 4, p. 1073, 2018, doi: 10.3390/s18041073.

[36] T. Á *et al.*, "Comparison between wireless and wired EEG recordings in a virtual reality lab: Case report," in *2014 5th IEEE Conference on Cognitive Infocommunications (CogInfoCom)*, 5-7 Nov. 2014 2014, pp. 599-603, doi: 10.1109/CogInfoCom.2014.7020414.

[37] A. Eskandarian, C. Wu, and C. Sun, "Research Advances and Challenges of Autonomous and Connected Ground Vehicles," *IEEE Transactions on Intelligent Transportation Systems,* pp. 1-29, 2019, doi: 10.1109/TITS.2019.2958352.

[38] D. Barwick, "Clinical Electroencephalography and Topographic Brain Mapping Technology and Practice," (in eng), *J Neurol Neurosurg Psychiatry,* vol. 52, no. 11, pp. 1322-1323, 1989. [Online]. Available: https://www.ncbi.nlm.nih.gov/pmc/articles/PMC1031680/.

[39] L. John, *Driver Distraction and Inattention*, 1 ed. (Advances in Research and Countermeasures). London, 2013, p. 464.

[40] T.-H. Nguyen and W.-Y. Chung, "Detection of Driver Braking Intention Using EEG Signals During Simulated Driving," (in eng), *Sensors (Basel),* vol. 19, no. 13, p. 2863, 2019, doi: 10.3390/s19132863.

[41] Y. W. Jing Wang, Hao Qu, Guanghua Xu, "EEG-based Fatigue Driving Detection Using Correlation Dimension," *JOurnal of Vibroengineering,* vol. 16, no. 1, p. 7, 2014.

[42] R. Foong, K. K. Ang, Z. Zhang, and C. Quek, "An iterative cross-subject negative-unlabeled learning algorithm for quantifying passive fatigue," *Journal of Neural Engineering,* vol. 16, no. 5, p. 056013, 2019/08/12 2019, doi: 10.1088/1741-2552/ab255d.

[43] R. Foong, K. K. Ang, and C. Quek, "Correlation of reaction time and EEG log bandpower from dry frontal electrodes in a passive fatigue driving simulation experiment," in *2017 39th Annual International Conference of the IEEE Engineering in Medicine and Biology Society (EMBC)*, 11-15 July 2017 2017, pp. 2482-2485, doi: 10.1109/EMBC.2017.8037360.

[44] M. Sazgar and M. G. Young, "EEG Artifacts," in *Absolute Epilepsy and EEG Rotation Review: Essentials for Trainees*. Cham: Springer International Publishing, 2019, pp. 149-162.

[45] X. Jiang, G.-B. Bian, and Z. Tian, "Removal of Artifacts from EEG Signals: A Review," (in eng), *Sensors (Basel),* vol. 19, no. 5, p. 987, 2019, doi: 10.3390/s19050987.

[46] R. N. Vigário, "Extraction of ocular artefacts from EEG using independent component analysis," *Electroencephalography and Clinical Neurophysiology,* vol. 103, no. 3, pp. 395-404, 1997/09/01/ 1997, doi: https://doi.org/10.1016/S0013-4694(97)00042-8.

[47] S. Hu, G. Zheng, and B. Peters, "Driver fatigue detection from electroencephalogram spectrum after electrooculography artefact removal," *IET Intelligent Transport Systems,* vol. 7, no. 1, pp. 105-113, 2013, doi: 10.1049/iet-its.2012.0045.

[48] M. T. Akhtar, W. Mitsuhashi, and C. J. James, "Employing spatially constrained ICA and wavelet denoising, for automatic removal of artifacts from multichannel EEG data," *Signal Processing,* vol. 92, no. 2, pp. 401-416, 2012/02/01/ 2012, doi: https://doi.org/10.1016/j.sigpro.2011.08.005.

[49] S. Barua, M. U. Ahmed, C. Ahlstrom, S. Begum, and P. Funk, "Automated EEG Artifact Handling With Application in Driver Monitoring," *IEEE Journal of Biomedical and Health Informatics,* vol. 22, no. 5, pp. 1350-1361, 2018, doi: 10.1109/JBHI.2017.2773999.

[50] I. Winkler, S. Haufe, and M. Tangermann, "Automatic Classification of Artifactual ICA-Components for Artifact Removal in EEG Signals," *Behavioral and Brain Functions,* vol. 7, no. 1, p. 30, 2011/08/02 2011, doi: 10.1186/1744-9081-7-30.

[51] W. D. Clercq, A. Vergult, B. Vanrumste, W. V. Paesschen, and S. V. Huffel, "Canonical Correlation Analysis Applied to Remove Muscle Artifacts From the Electroencephalogram," *IEEE Transactions on Biomedical Engineering,* vol. 53, no. 12, pp. 2583-2587, 2006, doi: 10.1109/TBME.2006.879459.

[52] A. S. Janani *et al.*, "Improved artefact removal from EEG using Canonical Correlation Analysis and spectral slope," *Journal of Neuroscience Methods,* vol. 298, pp. 1-15, 2018/03/15/ 2018, doi: https://doi.org/10.1016/j.jneumeth.2018.01.004.

[53] M. H. Soomro, N. Badruddin, M. Z. Yusoff, and M. A. Jatoi, "Automatic eye-blink artifact removal method based on EMD-CCA," in *2013 ICME International Conference on Complex Medical Engineering*, 25-28 May 2013 2013, pp. 186-190, doi: 10.1109/ICCME.2013.6548236.

[54] J. Gao, C. Zheng, and P. Wang, "Online Removal of Muscle Artifact from Electroencephalogram Signals Based on Canonical Correlation Analysis," *Clinical EEG and Neuroscience,* vol. 41, no. 1, pp. 53-59, 2010/01/01 2010, doi: 10.1177/155005941004100111.

[55] C.-T. Lin, C.-S. Huang, W.-Y. Yang, A. K. Singh, C.-H. Chuang, and Y.-K. Wang, "Real-Time EEG Signal Enhancement Using Canonical Correlation Analysis and Gaussian Mixture Clustering," (in eng), *J Healthc Eng,* vol. 2018, pp. 5081258-5081258, 2018, doi: 10.1155/2018/5081258.

[56] I. MathWorks, MATLAB : the language of technical computing : computation, visualization, programming : installation guide for UNIX version 5. Natwick : Math Works Inc., 1996., 1996.

[57] L. Buitinck *et al.*, "API design for machine learning software: experiences from the scikit-learn project," *ArXiv,* vol. abs/1309.0238, 2013.

[58] R. Polikar. "The Engineering Ultimate Guide to Wavelet Analysis: The Wavelet Tutorial." http://users.rowan.edu/~polikar/WTtutorial.html (accessed.

[59] S. Khatun, R. Mahajan, and B. I. Morshed, "Comparative Study of Wavelet-Based Unsupervised Ocular Artifact Removal Techniques for Single-Channel EEG Data," *IEEE Journal of Translational Engineering in Health and Medicine,* vol. 4, pp. 1-8, 2016, doi: 10.1109/JTEHM.2016.2544298.

[60] V. Krishnaveni, S. Jayaraman, S. Aravind, V. Hariharasudhan, and R. Kalidoss, "Automatic Identification and Removal of Ocular Artifacts from EEG using Wavelet Transform," *Meas. Sci. Rev.,* vol. 6, 11/30 2005.

[61] J. C. Woestenburg, M. N. Verbaten, and J. L. Slangen, "The removal of the eye-movement artifact from the EEG by regression analysis in the frequency domain," *Biological Psychology,* vol. 16, no. 1, pp. 127-147, 1983/02/01/ 1983, doi: https://doi.org/10.1016/0301-0511(83)90059-5.

[62] J. L. Kenemans, P. C. Molenaar, M. N. Verbaten, and J. L. Slangen, "Removal of the ocular artifact from the EEG: A comparison of time and frequency domain methods with simulated and real data," vol. 28, ed. United Kingdom: Blackwell Publishing, 1991, pp. 114-121.

[63] M. M. N. Mannan, M. A. Kamran, S. Kang, and M. Y. Jeong, "Effect of EOG Signal Filtering on the Removal of Ocular Artifacts and EEG-Based Brain-Computer Interface: A Comprehensive Study," *Complexity,* vol. 2018, p. 4853741, 2018/07/04 2018, doi: 10.1155/2018/4853741.

[64] M. F. Issa and Z. Juhasz, "Improved EOG Artifact Removal Using Wavelet Enhanced Independent Component Analysis," (in eng), *Brain Sci,* vol. 9, no. 12, p. 355, 2019, doi: 10.3390/brainsci9120355.

[65] M. M. N. Mannan, M. Y. Jeong, and M. A. Kamran, "Hybrid ICA-Regression: Automatic Identification and Removal of Ocular Artifacts from Electroencephalographic Signals," (in eng), *Frontiers in human*

[66] Z. M. Hira and D. F. Gillies, "A Review of Feature Selection and Feature Extraction Methods Applied on Microarray Data," (in eng), *Adv Bioinformatics,* vol. 2015, pp. 198363-198363, 2015, doi: 10.1155/2015/198363.

[67] S. Sur and V. K. Sinha, "Event-related potential: An overview," (in eng), *Ind Psychiatry J,* vol. 18, no. 1, pp. 70-73, 2009, doi: 10.4103/0972-6748.57865.

[68] G. Pfurtscheller, "Functional brain imaging based on ERD/ERS," *Vision Research,* vol. 41, no. 10, pp. 1257-1260, 2001/05/01/ 2001, doi: https://doi.org/10.1016/S0042-6989(00)00235-2.

[69] Giovanni, T. Suprihadi, and K. Karyono, "DROWTION: Driver drowsiness detection software using MINDWAVE," in *2014 International Conference on Industrial Automation, Information and Communications Technology*, 28-30 Aug. 2014 2014, pp. 141-144, doi: 10.1109/IAICT.2014.6922096.

[70] F. E. Bloom, "Chapter 1 - Fundamentals of Neuroscience," in *Fundamental Neuroscience (Fourth Edition)*, L. R. Squire, D. Berg, F. E. Bloom, S. du Lac, A. Ghosh, and N. C. Spitzer Eds. San Diego: Academic Press, 2013, pp. 3-13.

[71] *The cognitive neurosciences, 5th ed* (The cognitive neurosciences, 5th ed.). Cambridge, MA, US: MIT Press, 2014, pp. xvi, 1106-xvi, 1106.

[72] Y. Wang, S. Gao, and X. Gao, "Common Spatial Pattern Method for Channel Selelction in Motor Imagery Based Brain-computer Interface," in *2005 IEEE Engineering in Medicine and Biology 27th Annual Conference*, 17-18 Jan. 2006 2005, pp. 5392-5395, doi: 10.1109/IEMBS.2005.1615701.

[73] J. Kim, I. Kim, S. Haufe, and S. Lee, "Brain-computer interface for smart vehicle: Detection of braking intention during simulated driving," in *2014 International Winter Workshop on Brain-Computer Interface (BCI)*, 17-19 Feb. 2014 2014, pp. 1-3, doi: 10.1109/iww-BCI.2014.6782549.

[74] C. Dijksterhuis, D. de Waard, K. Brookhuis, B. Mulder, and R. de Jong, "Classifying visuomotor workload in a driving simulator using subject specific spatial brain patterns," (in English), *Front. Neurosci.,* Original Research vol. 7, no. 149, 2013-August-21 2013, doi: 10.3389/fnins.2013.00149.

[75] G. Dornhege, B. Blankertz, G. Curio, and K.-R. Müller, "Increase information transfer rates in BCI by CSP extension to multi-class," presented at the Proceedings of the 16th International Conference on Neural Information Processing Systems, Whistler, British Columbia, Canada, 2003.

[76] W. Wei, G. Xiaorong, and G. Shangkai, "One-Versus-the-Rest(OVR) Algorithm: An Extension of Common Spatial Patterns(CSP) Algorithm to Multi-class Case," in *2005 IEEE Engineering in Medicine and Biology 27th Annual Conference*, 17-18 Jan. 2006 2005, pp. 2387-2390, doi: 10.1109/IEMBS.2005.1616947.

[77] M. Grosse-Wentrup* and M. Buss, "Multiclass Common Spatial Patterns and Information Theoretic Feature Extraction," *IEEE Transactions on Biomedical Engineering,* vol. 55, no. 8, pp. 1991-2000, 2008, doi: 10.1109/TBME.2008.921154.

[78] A. Kai Keng, C. Zheng Yang, Z. Haihong, and G. Cuntai, "Filter Bank Common Spatial Pattern (FBCSP) in Brain-Computer Interface," in *2008 IEEE International Joint Conference on Neural Networks (IEEE World Congress on Computational Intelligence)*, 1-8 June 2008 2008, pp. 2390-2397, doi: 10.1109/IJCNN.2008.4634130.

[79] X. Song and S.-C. Yoon, "Improving brain–computer interface classification using adaptive common spatial patterns," *Computers in Biology and Medicine,* vol. 61, pp. 150-160, 2015/06/01/ 2015, doi: https://doi.org/10.1016/j.compbiomed.2015.03.023.

[80] [80]A. P. Costa, J. S. Møller, H. K. Iversen, and S. Puthusserypady, "An adaptive CSP filter to investigate user independence in a 3-class MI-BCI paradigm," *Computers in Biology and Medicine,* vol. 103, pp. 24-33, 2018/12/01/ 2018, doi: https://doi.org/10.1016/j.compbiomed.2018.09.021.

[81] [81]H. Yu, H. Lu, S. Wang, K. Xia, Y. Jiang, and P. Qian, "A General Common Spatial Patterns for EEG Analysis With Applications to Vigilance Detection," *IEEE Access,* vol. 7, pp. 111102-111114, 2019, doi: 10.1109/ACCESS.2019.2934519.

[82] [82]M. X. Cohen, "Analyzing neural time series data : theory and practice," (in English.), 2014. [Online]. Available: http://site.ebrary.com/id/10829847.

[83] [83]T. J. Y. K. Wang, C. Lin, "EEG-Based Attention Tracking During Distracted Driving," *IEEE Transactions on Neural Systems and Rehabilitation Engineering,* vol. 23, no. 6, 2015.

[84] T. Oliphant, *Guide to NumPy*. 2006.

[85] A. Y. Kaplan, A. A. Fingelkurts, A. A. Fingelkurts, S. V. Borisov, and B. S. Darkhovsky, "Nonstationary nature of the brain activity as revealed by EEG/MEG: Methodological, practical and conceptual challenges," *Signal Processing,* vol. 85, no. 11, pp. 2190-2212, 2005/11/01/ 2005, doi: https://doi.org/10.1016/j.sigpro.2005.07.010.

[86] H. S. AlZu'bi, W. Al-Nuaimy, and N. S. Al-Zubi, "EEG-based Driver Fatigue Detection," in *2013 Sixth International Conference on Developments in eSystems Engineering*, 16-18 Dec. 2013 2013, pp. 111-114, doi: 10.1109/DeSE.2013.28.

[87] F. Mohamed, S. F. Ahmed, Z. Ibrahim, and S. Yaacob, "Comparison of Features Based on Spectral Estimation for the Analysis of EEG Signals in Driver Behavior," in *2018 International Conference on Computational Approach in Smart Systems Design and Applications (ICASSDA)*, 15-17 Aug. 2018 2018, pp. 1-7, doi: 10.1109/ICASSDA.2018.8477633.

[88] C.-T. Lin, S.-A. Chen, T.-T. Chiu, H.-Z. Lin, and L.-W. Ko, "Spatial and temporal EEG dynamics of dual-task driving performance," (in eng), *Journal of neuroengineering and rehabilitation,* vol. 8, pp. 11-11, 2011, doi: 10.1186/1743-0003-8-11.

[89] M. Awais, N. Badruddin, and M. Drieberg, "EEG Brain Connectivity Analysis to Detect Driver Drowsiness Using Coherence," in *2017 International Conference on Frontiers of Information Technology (FIT)*, 18-20 Dec. 2017 2017, pp. 110-114, doi: 10.1109/FIT.2017.00027.

[90] S. Jung, H. Shin, and W. Chung, "Driver fatigue and drowsiness monitoring system with embedded electrocardiogram sensor on steering wheel," *IET Intelligent Transport Systems,* vol. 8, no. 1, pp. 43-50, 2014, doi: 10.1049/iet-its.2012.0032.

[91] C. Dumitrescu, I. M. Costea, F. Nemtanu, I. Badescu, and A. Banica, "Developing a multi sensors system to detect sleepiness to drivers from transport systems," in *2016 IEEE 22nd International Symposium for Design and Technology in Electronic Packaging (SIITME)*, 20-23 Oct. 2016 2016, pp. 175-178, doi: 10.1109/SIITME.2016.7777271.

[92] M. Apoorva, "Power Spectrum Density Estimation Methods for Michelson Interferometer Wavemters," Master of Applied Science, Electrical and Computer Engineering, University of Ottawa, 2016.

[93] J. Cecconi and J. Cecconi, "Spectral Analysis," (in English.), 2011. [Online]. Available: https://public.ebookcentral.proquest.com/choice/publicfullrecord.aspx?p=763580.

[94] R. Chai *et al.*, "Driver Fatigue Classification With Independent Component by Entropy Rate Bound Minimization Analysis in an EEG-Based System," *IEEE Journal of Biomedical and Health Informatics,* vol. 21, no. 3, pp. 715-724, 2017, doi: 10.1109/JBHI.2016.2532354.

[95] Z. Guo, Y. Pan, G. Zhao, S. Cao, and J. Zhang, "Detection of Driver Vigilance Level Using EEG Signals and Driving Contexts," *IEEE Transactions on Reliability,* vol. 67, no. 1, pp. 370-380, 2018, doi: 10.1109/TR.2017.2778754.

[96] A. Picot, S. Charbonnier, and A. Caplier, "On-Line Detection of Drowsiness Using Brain and Visual Information," *IEEE Transactions on Systems, Man, and Cybernetics - Part A: Systems and Humans,* vol. 42, no. 3, pp. 764-775, 2012, doi: 10.1109/TSMCA.2011.2164242.

[97] H. Xue-Qin, W. Zheng, and B. Lu, "Driving fatigue detection with fusion of EEG and forehead EOG," in *2016 International Joint Conference on Neural Networks (IJCNN)*, 24-29 July 2016 2016, pp. 897-904, doi: 10.1109/IJCNN.2016.7727294.

[98] V. Alizadeh and O. Dehzangi, "The impact of secondary tasks on drivers during naturalistic driving: Analysis of EEG dynamics," in *2016 IEEE 19th International Conference on Intelligent Transportation Systems (ITSC)*, 1-4 Nov. 2016 2016, pp. 2493-2499, doi: 10.1109/ITSC.2016.7795957.

[99] A. G. Correa and E. L. Leber, "An automatic detector of drowsiness based on spectral analysis and wavelet decomposition of EEG records," in *2010 Annual International Conference of the IEEE Engineering in Medicine and Biology*, 31 Aug.-4 Sept. 2010 2010, pp. 1405-1408, doi: 10.1109/IEMBS.2010.5626721.

[100] B. P. Nayak, S. Kar, A. Routray, and A. K. Padhi, "A biomedical approach to retrieve information on driver's fatigue by integrating

[100] EEG, ECG and blood biomarkers during simulated driving session," in *2012 4th International Conference on Intelligent Human Computer Interaction (IHCI)*, 27-29 Dec. 2012 2012, pp. 1-6, doi: 10.1109/IHCI.2012.6481812.

[101] A. Sengupta, A. Routray, and S. Datta, "Brain networks using nonlinear interdependence-based EEG synchronization: A study of human fatigue," in *2016 International Conference on Systems in Medicine and Biology (ICSMB)*, 4-7 Jan. 2016 2016, pp. 170-173, doi: 10.1109/ICSMB.2016.7915114.

[102] M. Mohammadpour and S. Mozaffari, "Classification of EEG-based attention for brain computer interface," in *2017 3rd Iranian Conference on Intelligent Systems and Signal Processing (ICSPIS)*, 20-21 Dec. 2017 2017, pp. 34-37, doi: 10.1109/ICSPIS.2017.8311585.

[103] R. N. Khushaba, S. Kodagoda, S. Lal, and G. Dissanayake, "Driver Drowsiness Classification Using Fuzzy Wavelet-Packet-Based Feature-Extraction Algorithm," *IEEE Transactions on Biomedical Engineering*, vol. 58, no. 1, pp. 121-131, 2011, doi: 10.1109/TBME.2010.2077291.

[104] C. Zhang, F. Cong, and H. Wang, "Driver fatigue analysis based on binary brain networks," in *2017 Seventh International Conference on Information Science and Technology (ICIST)*, 16-19 April 2017 2017, pp. 485-489, doi: 10.1109/ICIST.2017.7926809.

[105] A. Saha, A. Konar, and A. K. Nagar, "EEG Analysis for Cognitive Failure Detection in Driving Using Type-2 Fuzzy Classifiers," *IEEE Transactions on Emerging Topics in Computational Intelligence*, vol. 1, no. 6, pp. 437-453, 2017, doi: 10.1109/TETCI.2017.2750761.

[106] C. Lin *et al.*, "Classification of Driver's Cognitive Responses Using Nonparametric Single-trial EEG Analysis," in *2007 IEEE International Symposium on Circuits and Systems*, 27-30 May 2007 2007, pp. 2019-2023, doi: 10.1109/ISCAS.2007.378434.

[107] C.-T. Lin, K.-L. Lin, L.-W. Ko, S.-F. Liang, B.-C. Kuo, and I. F. Chung, "Nonparametric Single-Trial EEG Feature Extraction and Classification of Driver's Cognitive Responses," *EURASIP Journal on Advances in Signal Processing*, vol. 2008, no. 1, p. 849040, 2008/03/31 2008, doi: 10.1155/2008/849040.

[108] C. E. Shannon, "A mathematical theory of communication," *The Bell System Technical Journal*, vol. 27, no. 3, pp. 379-423, 1948, doi: 10.1002/j.1538-7305.1948.tb01338.x.

[109] J. M. Yentes, N. Hunt, K. K. Schmid, J. P. Kaipust, D. McGrath, and N. Stergiou, "The appropriate use of approximate entropy and sample entropy with short data sets," *Ann Biomed Eng*, vol. 41, no. 2, pp. 349-365, 2013, doi: 10.1007/s10439-012-0668-3.

[110] U. Budak, V. Bajaj, Y. Akbulut, O. Atila, and A. Sengur, "An Effective Hybrid Model for EEG-Based Drowsiness Detection," *IEEE Sensors Journal*, vol. 19, no. 17, pp. 7624-7631, 2019, doi: 10.1109/JSEN.2019.2917850.

[111] Z. Gao *et al.*, "Relative Wavelet Entropy Complex Network for Improving EEG-Based Fatigue Driving Classification," *IEEE Transactions on Instrumentation and Measurement*, vol. 68, no. 7, pp. 2491-2497, 2019, doi: 10.1109/TIM.2018.2865842.

[112] J. Hu, F. Liu, and P. Wang, "EEG-Based Multiple Entropy Analysis for Assessing Driver Fatigue," in *2019 5th International Conference on Transportation Information and Safety (ICTIS)*, 14-17 July 2019 2019, pp. 1290-1294, doi: 10.1109/ICTIS.2019.8883591.

[113] A. Chaudhuri and A. Routray, "Driver Fatigue Detection Through Chaotic Entropy Analysis of Cortical Sources Obtained From Scalp EEG Signals," *IEEE Transactions on Intelligent Transportation Systems*, pp. 1-14, 2019, doi: 10.1109/TITS.2018.2890332.

[114] Z. Khaliliardali, R. Chavarriaga, L. A. Gheorghe, and J. d. R. Millán, "Detection of anticipatory brain potentials during car driving," in *2012 Annual International Conference of the IEEE Engineering in Medicine and Biology Society*, 28 Aug.-1 Sept. 2012 2012, pp. 3829-3832, doi: 10.1109/EMBC.2012.6346802.

[115] Z. Y. Chin, X. Zhang, C. Wang, and K. K. Ang, "EEG-based discrimination of different cognitive workload levels from mental arithmetic," in *2018 40th Annual International Conference of the IEEE Engineering in Medicine and Biology Society (EMBC)*, 18-21 July 2018 2018, pp. 1984-1987, doi: 10.1109/EMBC.2018.8512675.

[116] T. Teng, L. Bi, and X. Fan, "Using EEG to recognize emergency situations for brain-controlled vehicles," in *2015 IEEE Intelligent Vehicles Symposium (IV)*, 28 June-1 July 2015 2015, pp. 1305-1309, doi: 10.1109/IVS.2015.7225896.

[117] L. Bi, J. Zhang, and J. Lian, "EEG-Based Adaptive Driver-Vehicle Interface Using Variational Autoencoder and PI-TSVM," *IEEE Transactions on Neural Systems and Rehabilitation Engineering*, vol. 27, no. 10, pp. 2025-2033, 2019, doi: 10.1109/TNSRE.2019.2940046.

[118] C.-C. Chang and C.-J. Lin, "LIBSVM: A library for support vector machines," *ACM Trans. Intell. Syst. Technol.*, vol. 2, no. 3, p. Article 27, 2011, doi: 10.1145/1961189.1961199.

[119] J. Hu, "Automated Detection of Driver Fatigue Based on AdaBoost Classifier with EEG Signals," (in eng), *Frontiers in computational neuroscience*, vol. 11, pp. 72-72, 2017, doi: 10.3389/fncom.2017.00072.

[120] L.-Y. Hu, M.-W. Huang, S.-W. Ke, and C.-F. Tsai, "The distance function effect on k-nearest neighbor classification for medical datasets," (in eng), *Springerplus*, vol. 5, no. 1, pp. 1304-1304, 2016, doi: 10.1186/s40064-016-2941-7.

[121] N. A. B. Amirudin, N. Saad, S. S. A. Ali, and S. H. Adil, "Detection and Analysis of Driver Drowsiness," in *2018 3rd International Conference on Emerging Trends in Engineering, Sciences and Technology (ICEEST)*, 21-22 Dec. 2018 2018, pp. 1-9, doi: 10.1109/ICEEST.2018.8643326.

[122] T. L. Fine, "Feedforward neural network methodology," (in English), 2005. [Online]. Available: http://accesbib.uqam.ca/cgi-bin/bduqam/transit.pl?&noMan=25126728.

[123] R. A. Sayed, A. Eskandarian, and M. Oskard, "DRIVER DROWSINESS DETECTION USING ARTIFICIAL NEURAL NETWORKS," 2001.

[124] C. Zhang, H. Wang, and R. Fu, "Automated Detection of Driver Fatigue Based on Entropy and Complexity Measures," *IEEE Transactions on Intelligent Transportation Systems*, vol. 15, no. 1, pp. 168-177, 2014, doi: 10.1109/TITS.2013.2275192.

[125] S.-C. Ng, C.-C. Cheung, A. k.-f. Lui, and S. Xu, "Magnified Gradient Function to Improve First-Order Gradient-Based Learning Algorithms," in *Advances in Neural Networks – ISNN 2012*, Berlin, Heidelberg, J. Wang, G. G. Yen, and M. M. Polycarpou, Eds., 2012// 2012: Springer Berlin Heidelberg, pp. 448-457.

[126] L. M. King, H. T. Nguyen, and S. K. L. Lal, "Early Driver Fatigue Detection from Electroencephalography Signals using Artificial Neural Networks," in *2006 International Conference of the IEEE Engineering in Medicine and Biology Society*, 30 Aug.-3 Sept. 2006 2006, pp. 2187-2190, doi: 10.1109/IEMBS.2006.259231.

[127] M. A. Almogbel, A. H. Dang, and W. Kameyama, "EEG-signals based cognitive workload detection of vehicle driver using deep learning," in *2018 20th International Conference on Advanced Communication Technology (ICACT)*, 11-14 Feb. 2018 2018, pp. 256-259, doi: 10.23919/ICACT.2018.8323716.

[128] M. Hajinoroozi, Z. Mao, T.-P. Jung, C.-T. Lin, and Y. Huang, "EEG-based prediction of driver's cognitive performance by deep convolutional neural network," *Signal Processing: Image Communication*, vol. 47, pp. 549-555, 2016/09/01/ 2016, doi: https://doi.org/10.1016/j.image.2016.05.018.

[129] Z. Gao *et al.*, "EEG-Based Spatio–Temporal Convolutional Neural Network for Driver Fatigue Evaluation," *IEEE Transactions on Neural Networks and Learning Systems*, vol. 30, no. 9, pp. 2755-2763, 2019, doi: 10.1109/TNNLS.2018.2886414.

[130] M. Hajinoroozi, J. M. Zhang, and Y. Huang, "Driver's fatigue prediction by deep covariance learning from EEG," in *2017 IEEE International Conference on Systems, Man, and Cybernetics (SMC)*, 5-8 Oct. 2017 2017, pp. 240-245, doi: 10.1109/SMC.2017.8122609.

[131] M. Cogswell, F. Ahmed, R. B. Girshick, C. L. Zitnick, and D. Batra, "Reducing Overfitting in Deep Networks by Decorrelating Representations," *CoRR*, vol. abs/1511.06068, 2015.

[132] D. P. Mandic, J. Chambers, *"Fundamentals,"* in *Recurrent Neural Networks for Prediction* (1st ed.), pp. 9-29.

[133] M. Moinnereau, S. Karimian-Azari, T. Sakuma, H. Boutani, L. Gheorghe, and T. H. Falk, "EEG Artifact Removal for Improved Automated Lane Change Detection while Driving," in *2018 IEEE International Conference on Systems, Man, and Cybernetics (SMC)*, 7-10 Oct. 2018 2018, pp. 1076-1080, doi: 10.1109/SMC.2018.00190.

[134] S. Lee, J. Kim, and S. Lee, "Detecting Driver's Braking Intention Using Recurrent Convolutional Neural Networks Based EEG Analysis," in *2017 4th IAPR Asian Conference on Pattern Recognition (ACPR)*, 26-29 Nov. 2017 2017, pp. 840-845, doi: 10.1109/ACPR.2017.86.

[135] Y. Liu, Y. Lin, S. Wu, C. Chuang, and C. Lin, "Brain Dynamics in Predicting Driving Fatigue Using a Recurrent Self-Evolving Fuzzy Neural Network," *IEEE Transactions on Neural Networks and*

*Learning Systems,* vol. 27, no. 2, pp. 347-360, 2016, doi: 10.1109/TNNLS.2015.2496330.
[136] A. Saha, A. Konar, R. Burman, and A. K. Nagar, "EEG analysis for cognitive failure detection in driving using neuro-evolutionary synergism," in *2014 International Joint Conference on Neural Networks (IJCNN)*, 6-11 July 2014 2014, pp. 2108-2115, doi: 10.1109/IJCNN.2014.6889929.
[137] A. Konar, Computational intellingence : principles, techniques, and applications. New Delhi (India): Springer (in English), 2007.
[138] C. Lin *et al.*, "Adaptive EEG-Based Alertness Estimation System by Using ICA-Based Fuzzy Neural Networks," *IEEE Transactions on Circuits and Systems I: Regular Papers,* vol. 53, no. 11, pp. 2469-2476, 2006, doi: 10.1109/TCSI.2006.884408.
[139] Y. Liu, Y. Lin, S. Wu, C. Chuang, M. Prasad, and C. Lin, "EEG-based driving fatigue prediction system using functional-link-based fuzzy neural network," in *2014 International Joint Conference on Neural Networks (IJCNN)*, 6-11 July 2014 2014, pp. 4109-4113, doi: 10.1109/IJCNN.2014.6889736.
[140] Y. Liu *et al.*, "Driving fatigue prediction with pre-event electroencephalography (EEG) via a recurrent fuzzy neural network," in *2016 IEEE International Conference on Fuzzy Systems (FUZZ-IEEE)*, 24-29 July 2016 2016, pp. 2488-2494, doi: 10.1109/FUZZ-IEEE.2016.7738006.
[141] C. Lin, I. Chung, L. Ko, Y. Chen, S. Liang, and J. Duann, "EEG-Based Assessment of Driver Cognitive Responses in a Dynamic Virtual-Reality Driving Environment," *IEEE Transactions on Biomedical Engineering,* vol. 54, no. 7, pp. 1349-1352, 2007, doi: 10.1109/TBME.2007.891164.
[142] C. Lin, S. Tsai, and L. Ko, "EEG-Based Learning System for Online Motion Sickness Level Estimation in a Dynamic Vehicle Environment," *IEEE Transactions on Neural Networks and Learning Systems,* vol. 24, no. 10, pp. 1689-1700, 2013, doi: 10.1109/TNNLS.2013.2275003.